\documentclass[ACS,STIX1COL]{WileyNJD-v2}
\articletype{Article Type}%
\usepackage{graphicx,multirow,array,amsmath,empheq,amsthm,amssymb,amsfonts,tikz,ragged2e,color,graphicx,siunitx,subcaption,float,adjustbox}
\begin{document}
\title{Damped dust-ion-acoustic solitons in collisional magnetized nonthermal plasmas}
\author[1]{M. R. Hassan$^{1}$}
\author[1]{S. Sultana$^{1}$}
\authormark{M. R. Hassan \textsc{et al}}
\address{$^1$ Department of Physics, Jahangirnagar University, Savar, Dhaka-1342, Bangladesh}
\corres{M. R. Hassan\\ Department of Physics, Jahangirnagar University, Savar, Dhaka-1342, Bangladesh\\
\email{hassan148phy@gmail.com}}
\presentaddress{Department of Physics, Jahangirnagar University, Savar, Dhaka-1342, Bangladesh}
\abstract{A multi-species magnetized collisional nonthermal plasma system, containing inertial ion species, non-inertial electron species following nonthermal $\kappa-$ distribution, and immobile dust particles, is considered to examine the characteristics of the dissipative dust-ion-acoustic (DIA) soliton modes, \textbf{theoretically and parametrically}. The electrostatic solitary modes are found to be associated with the low frequency dissipative dust-ion-acoustic solitary waves (DIASWs). The ion-neutral collision is taken into account, and the influence of ion-neutral collisional effects on the dynamics of dissipative DIASWs is investigated. It is reported that most of the plasma medium in space and laboratory are far from thermal equilibrium, and the particles in such plasma system are well fitted via the $\kappa-$nonthermal distribution than via the thermal Maxwellian distribution. The reductive perturbation approach is adopted to derive the damped KdV (dKdV) equation, and the solitary wave solution of the dKdV equation is derived via the tangent hyperbolic method to analyze the basic features (amplitude, width, speed, time evolution, etc.) of dissipative DIASWs. The propagation nature and also the basic features of dissipative DIASWs are seen to influence significantly due to the variation of the plasma configuration parameters and also due to the variation of the supethermality index $\kappa$ in the considered plasma system. The implication of the results of this study could be useful for better understanding the electrostatic localized disturbances, in the ion length and time scale, in space and experimental dusty plasmas, where the presence of excess energetic electrons and ion-neutral collisional damping are accountable.}
\maketitle
\footnotetext{}
\section{Introduction}
\label{1sec:Introduction}
The origin and propagation of nonlinear electrostatic waves in dusty plasma have drawn the attention of researchers in the last few decades after the first prediction of the existence of these waves in Saturn's ring by Bliokh and Yarashenko in 1985 \cite{Bliokh1985}, after which scientists discovered several new modes of nonlinear excitations, e.g., dust ion acoustic waves (DIAWs) \cite{Shukla1992}, dust acoustic waves (DAWs) \cite{Rao1990}, dust lattice waves (DLWs) \cite{Piel2002}, etc. Dust ion acoustic solitary structures are a type of dust-ion-acoustic wave (DIAW) are formed in a plasma system due to the compromise between the dispersion and the nonlinearity and \cite{Naheer1978} form stable hump or dip shaped structures. In a DIAW, the mass density of ions is considered to provide the inertia, while the electrons thermal pressure is assumed to give the restoring force to generate the wave, and the negatively charged dust species are approximated to remain static. Viking Satellite \cite{Temerin1982}and THEMIS mission \cite{Ergun1998} have already identified this type of wave in both spaces (viz., solar atmosphere \cite{Panwar2014}, Saturn's magnetosphere \cite{Panwar2014}, pulsar magnetosphere \cite{Rehman2016}, active galactic nuclei \cite{Chowdhury2017}, neutron star's polar region \cite{Chowdhury2018}, etc.) and laboratory environments (viz., semiconductor plasmas \cite{Chowdhury2017}, cathode discharge \cite{Rehman2016}, tokamaks \cite{Chowdhury2017}, etc.).

From the rudimentary stage of plasma physics, it was established that only a fractional part of electrons acquire higher kinetic energy than that of the most remaining and thus was modeled to follow the Maxwell distribution \cite{Chappell1980}. After observing the dominating part of electrons possessing excess energy in space plasmas \cite{Vasyliunas1968,Armstrong1983,Fitzenreiter1998,Vocks2008} and laboratory environments \cite{Kunze1968,Ma1998,Yoon2005}, they were formulated to follow the superthermal $\kappa$-distribution \cite{Hasegawa1985,Mace1995,Hellberg2009}. The researchers have carried out much research work in the last few decades to study the origin, propagation, and characteristics of the rogue structures \cite{Abdelwahed2016,Sing2019,Alinejad2019}, shock structures \cite{Pakzad2011,Sultana2012,Chahal2017}, envelope solitons \cite{Noman2019,Sultana2011,Gill2010}, solitary structures \cite{Saini2009,Shahmansouri2013,Farooq2017,Sultana2015,Sultana2018}, etc. plasmas containing kappa-distributed electrons/ positrons/ ions.

There is always a chance for the solitary pulses to get damped due to dissipation in the plasma medium gradually. Apart from the existence of the collision between different plasma components \cite{Vladimirov1999,Krapak2001,Kashkari2021}, the ion-acoustic solitary structures can also get dissipated if the ions no longer remain cold instead become hot enough compared to that of the electron species, i.e., due to the ion temperature effect \cite{Mamun1997,Chatterjee2009,Roy2012,Mehdipoor2020}.  As most of the practical systems around us are not in equilibrium, almost every wave has to suffer a certain amount (more or less) of dissipation. This influenced the researchers to study solitary wave's attenuation in various plasma mediums \cite{Nicholson1976,Pereira1977,Dutta2012,Sultana2015} and nonlinear optics \cite{Mollenauer1980,Nozaki1983,Afanasjev1995,Soto1997}. Mamun \cite{Mamun1997} in 1997 considered a plasma medium containing no dust and studied the effect of ion temperature on electrostatic solitary structures in unmagnetized collisionless non-thermal plasmas. Later, Chatterjee et al. \cite{Chatterjee2009} carried out a similar investigation to determine ion temperature's effect on the solitary structures in a quantum electron-ion plasma. Roy et al. \cite{Roy2012} also conducted a research work analogous to Mamun in which he assumed a plasma consisting of a q-nonextensive electron species. Sing \cite{Sing2013} extended the work of Chatterjee by considering the influence of the magnetic field. Sultana \cite{Sultana2018} took a huge step to consider the magnetic field effect and included dust in her system, but she did not take the effect of hot ion fluid into account in her investigation. This motivated us to observe the origin, evaluation, streaming, and the fundamental properties of nonlinear solitary excitations in a plasma medium comprised of negatively charged dust, hot ion fluid colliding with neutral particles, and super-thermal electrons in the presence of an external magnetic field. The investigation of such a plasma system will enable us to analyze and understand the nonlinear structure's behavior in a magnetized complex dusty plasma in which the ion temperature effect and the energy dissipation effect are considered at the same time.

This paper is organized in the following manner: The basic governing equations for describing our plasma system are shown in section \ref{1sec:The Governing Equations}. Section \ref{1sec:Derivation of damped KdV equation} contains the derivation of the damped KdV equation using the reductive perturbation model. The analytical solutions representing dissipated solitary pulses are given in section \ref{1sec:ANALYTICAL SOLUTION}. The numerical simulations are shown in section \ref{1sec:Numerical simulation}. Finally, the summary of our research work is discussed briefly in section \ref{1sec:Discussion}.

\section{The plasma model and basic formalism}\label{1sec:The Governing Equations}
To study the nature of the low frequency obliquely propagating solitary waves in ion time (length) scale, we consider a three-component magnetized collisional complex/dusty plasma system consisting of

\begin{itemize}
\item  inertial ion fluid of mass $m_i$ and charge $z_ie$ with $e$ being the electronic charge and $z_i$ being the ion charge state;
\item inertialess electrons of mass $m_e$ and charge $-e$ following nonthermal $\kappa-$ distribution; and
\item micron/submicron sized massive negatively charged static dust of mass $m_d$ and charge $-z_de$ with $z_d$ being the dust charge state.
\end{itemize}
Thus, at equilibrium the plasma quasi-neutrality condition reads: $n_{e0} + z_dn_{d0} = z_in_{i0}$, where
$n_{s0}$ represents the number density of plasma species $s$ at equilibrium (here $s = e, i, d$ correspond to electron, ion, and
dust respectively). We also assume that the ambient constant magnetic field is acting along the $z$-axis (i.e.,
$\textbf{B}_0=B_0\hat{z}$). It is noted that the phase speed of DIAWs in such a plasma medium is much greater than the thermal
speed of ion and dust but much smaller than the electron's thermal speed (i.e., $v_{th,e}\gg v_{ph}\gg v_{th,i,d}$).
The ion-neutral collisions and also the thermal effect of inertial ion population on the dynamics of obliquely propagating DIASWs
are taken into account. One can, therefore, describe the dynamics of DIASWs in such a dusty plasma medium by the following
set of fluid equations

\begin{eqnarray}
&&\hspace*{-1.3cm}\frac{\partial n_i}{\partial T}+\nabla .(n_i\textbf{u}_i)=0,
\label{1:eq1}\\
&&\hspace*{-1.3cm}\frac{\partial \textbf{u}_i}{\partial T}+ (\textbf{u}_i.\nabla)\textbf{u}_i=-\frac{z_ie}{m_i}\nabla\Phi+\frac{z_ieB_0}{m_i}(\textbf{u}_i\times\hat{z})
\nonumber\\
&&\hspace*{1.85cm}-\frac{k_BT_i}{m_in_i}\nabla n_i-\upsilon_{in}\textbf{u}_i,
\label{1:eq2}\\
&&\hspace*{-1.3cm}\nabla^2\Phi=4\pi e(n_e+z_dn_{d0}-z_in_i),
\label{1:eq3}\
\end{eqnarray}
where $n_i$, $n_e$, and $n_{d0}$ are number densities of ions, electrons, and dust grains, respectively. $u_i$ is the velocity of inertial
ion fluid, $\Phi$ is the electrostatic potential, $T_i$ is the characteristic temperature of
ion fluid, and $\upsilon_{in}$ is the ion-neutral collision frequency.\textbf{It is worth mentioning that the collisional effect (due to charged particles and neutrals) can be neglected for cold plasma \cite{Sultana2021}, while this effect can be ignored for a hot plasma\cite{Mamun2021}. However, our aim here in this study is to see how collision affects the solitary waves in a warm plasma and we introduce a very small amount of collisional effect by replacing $\nu_{in}$ in terms of $\nu^{\prime}$, which will be explained in later section.}\\

The number density expression for the non-thermal $\kappa-$distributed electron has the form \cite{Sultana2012,Sultana2018,Thorne1991,Singla2021}
\begin{eqnarray}
&&\hspace*{-1.3cm}n_e=n_{e0}\left[1 -\frac{e \Phi}{k_BT_e(\kappa-3/2)}\right]^{-\kappa+\frac{1}{2}},
\label{1:eq4}\
\end{eqnarray}
here $T_e$ is the characteristic temperature of the electron, $k_B$ is the Boltzmann constant, and $\kappa$ is the
superthermality parameter which determines strength of superthermality or non-thermality of the plasma medium. It is
clear in (\ref{1:eq4}) and also in Refs. \cite{Maksimovic1997}, \cite{Hellberg2009}, and \cite{Panwar2014} that for a physically meaningful particle distribution,
one should consider $\kappa>3/2$. It should also be noted from existing research \cite{Abdelwahed2016,Sultana2012} that the values of $\kappa$
can lie in the range  $\infty\geq\kappa\geq 3/2$ and smaller values of $\kappa$ defines stronger non-thermality
and larger values of $\kappa$ means weaker non-thermality, and the plasma medium is considered to be Maxwellian
for the limit $\kappa\rightarrow\infty$.

To normalize our plasma model equations (\ref{1:eq1}) - (\ref{1:eq3}), we consider the following scaling factors
\begin{eqnarray}
&&\hspace*{-1.3cm}\nabla\rightarrow\frac{\nabla}{\lambda_D},\:t\rightarrow\frac{T}{\omega^{-1}_{pi}},\: u\rightarrow\frac{u_i}{C_i},\:n\rightarrow\frac{n_i}{n_{i0}},
\nonumber\\
&&\hspace*{-.45cm}\phi\rightarrow\frac{e\Phi}{k_BT_e},\nu_{in}\rightarrow\omega_{pi}\nu_i.
\nonumber\
\end{eqnarray}
The Debye length $\lambda_D$ and the characteristic plasma frequency $\omega_{pi}$ are given by
\begin{eqnarray}
\begin{rcases}
\lambda_D=\sqrt{\frac{k_BT_e}{4\pi n_{i0}z_ie^2}},\\ \omega_{pi}=\sqrt{\frac{4\pi n_{i0}z_i^2e^2}{m_i}},
\end{rcases}
\label{1:eq5}\
\end{eqnarray}
so that the condition of ion thermal speed $C_i=\omega_{pi}\lambda_D=(z_ik_BT_e/m_i)^{1/2}$ can be fulfilled.  Now if we take
$\mu=z_dn_{d0}/z_in_{nio}$, $\Omega_{ci}=\omega_{ci}/\omega_{pi}$ (in which $\omega_{ci}=z_ieB_0/m_i$), and $\sigma=T_i/(z_iT_e)$,
the normalized form of  equations \eqref{1:eq1}-\eqref{1:eq3} can be written as
\begin{eqnarray}
&&\hspace*{-1.3cm}\frac{\partial n}{\partial t}+\nabla .(n\textbf{u})=0,
\label{1:eq6}\\
&&\hspace*{-1.3cm}n\frac{\partial\textbf{u}}{\partial t}+n(\textbf{u}.\nabla)\textbf{u}=-n\nabla\phi+
\Omega_{ci}n(\textbf{u}\times\hat{z})
\nonumber\\
&&\hspace*{1.85cm}-\sigma\nabla n-n\nu_{i}\textbf{u},
\label{1:eq7}\\
&&\hspace*{-1.3cm}\nabla^2\phi=1-n+p\phi+q\phi^2+r\phi^3+\cdots.
\label{1:eq8}\
\end{eqnarray}

The coefficients $p$, $q$, and $q$ [appear in equation \eqref{1:eq8}], which contain the information of plasma non(super)thermality via the superthermality index $\kappa$, are expressed as
\begin{equation}
\begin{rcases}
p=\frac{(1-\mu)(\kappa-0.5)}{\kappa-1.5},\\q=\frac{(1-\mu)(\kappa-0.5)(\kappa+0.5)}{2(\kappa-1.5)^2},\\
r=\frac{(1-\mu)(\kappa-0.5)(\kappa+0.5)(\kappa+1.5)}{6(\kappa-1.5)^3}.
\end{rcases}
\label{1:eq9}\
\end{equation}

As mentioned above, for the  limit $\kappa\rightarrow\infty$, the distribution will no longer be nonthermal and the particle
distribution approaches to the thermal Maxwellian. We will consider the numerical value of superthermality index
$\kappa\geq3$ to study the nature of obliquely propagating DIASWs in magnetized collisional plasmas in the later
sections of this manuscript. Otherwise, the higher-order terms for the range $1.5<\kappa\leq3$  in Eq. \eqref{1:eq8} will become so significant that
neglecting them can cause to decrease the efficiency and the precision of the study to some extent.

\section{Damped DIASW\lowercase{s}: Perturbative approach}
\label{1sec:Derivation of damped KdV equation}
To examine the nonlinear dynamics of small but finite amplitude solitary waves in magnetized collisional plasmas, we introduce the stretched coordinates as
\begin{eqnarray}
\begin{rcases}
\xi=\epsilon^{1/2}(l_xx+l_yy+l_zz-Vt)),\\ \tau=\epsilon^{3/2}t,
\end{rcases}
\label{1:eq10}\
\end{eqnarray}
where $\epsilon$ is a small expansion parameter which characterizes the weakness of the dispersion
($0<\epsilon<1$) and $V$ is the phase speed of DIAWs (normalized by $C_i$) and to be determined later. Here  $l_x$, $1_y$, and $l_z$
symbolize the directional cosines of the wave vector $\textbf{k}$ along $x$, $y$, and $z$-axes, respectively to the magnetic field
$\textbf{B}_0$ so that $l_x^2+l_y^2+l_z^2=1$. As we considered weak damping in our
plasma medium due to ion-neutral collision, we can therefore assume the following scaling for the ion-neutral
collision frequency $\nu_i$ as
\begin{eqnarray}
&&\hspace*{-1.3cm}\nu_i=\epsilon^{3/2}\nu^{\prime},
\label{1:eq11}\
\end{eqnarray}
\textbf{where $\epsilon\ll1$ -- suggesting the presence of small amount of collisional effect in the considered plasma medium.} It is noted that the collisional frequency is extremely small in our medium.
Now, we expand the physical variables $n$, $\textbf{u}$, and $\phi$ near equilibrium in the power series of $\epsilon$ as
\begin{eqnarray}
\begin{pmatrix}n\\u_{x,y}\\u_z\\ \phi\end{pmatrix}=\begin{pmatrix}1\\0\\0\\0\end{pmatrix}+
\epsilon\begin{pmatrix} n_1\\ \epsilon^{1/2}u_{1x,y}\\ u_{1z}\\ \phi_1\end{pmatrix}
+\epsilon^2\begin{pmatrix}n_2\\u_{2x,y}\\u_{2z}\\ \phi_3\end{pmatrix}+\cdots
\label{1:eq12}\
\end{eqnarray}

Now, we substitute our assumptions in \eqref{1:eq10} -  \eqref{1:eq12} into the equations \eqref{1:eq6}-\eqref{1:eq8} and separating
the lowest order terms of $\epsilon$ from the resultant equations, we obtain the phase speed of obliquely propagating dust-
acoustic waves (DAWs) in the form
\begin{eqnarray}
&&\hspace*{-1.3cm}V=\sqrt{\frac{l_z^2}{p}+\sigma l_z^2}.
\label{1:eq13}\
\end{eqnarray}

It is seen from Eq. \eqref{1:eq13} that the phase speed of DIAWs depends on the obliquity angle $\delta$
($=\textmd{cos}^{-1}l_z$), the number densities of dust and ions (via the dust-to-ion number density $\mu$), the ion
and electron temperature (via $\sigma=T_i/T_e$)a ratio of ion, and the superthermality index $\kappa$.

\begin{figure}[!h]
\begin{center}
\includegraphics[width=6.7cm]{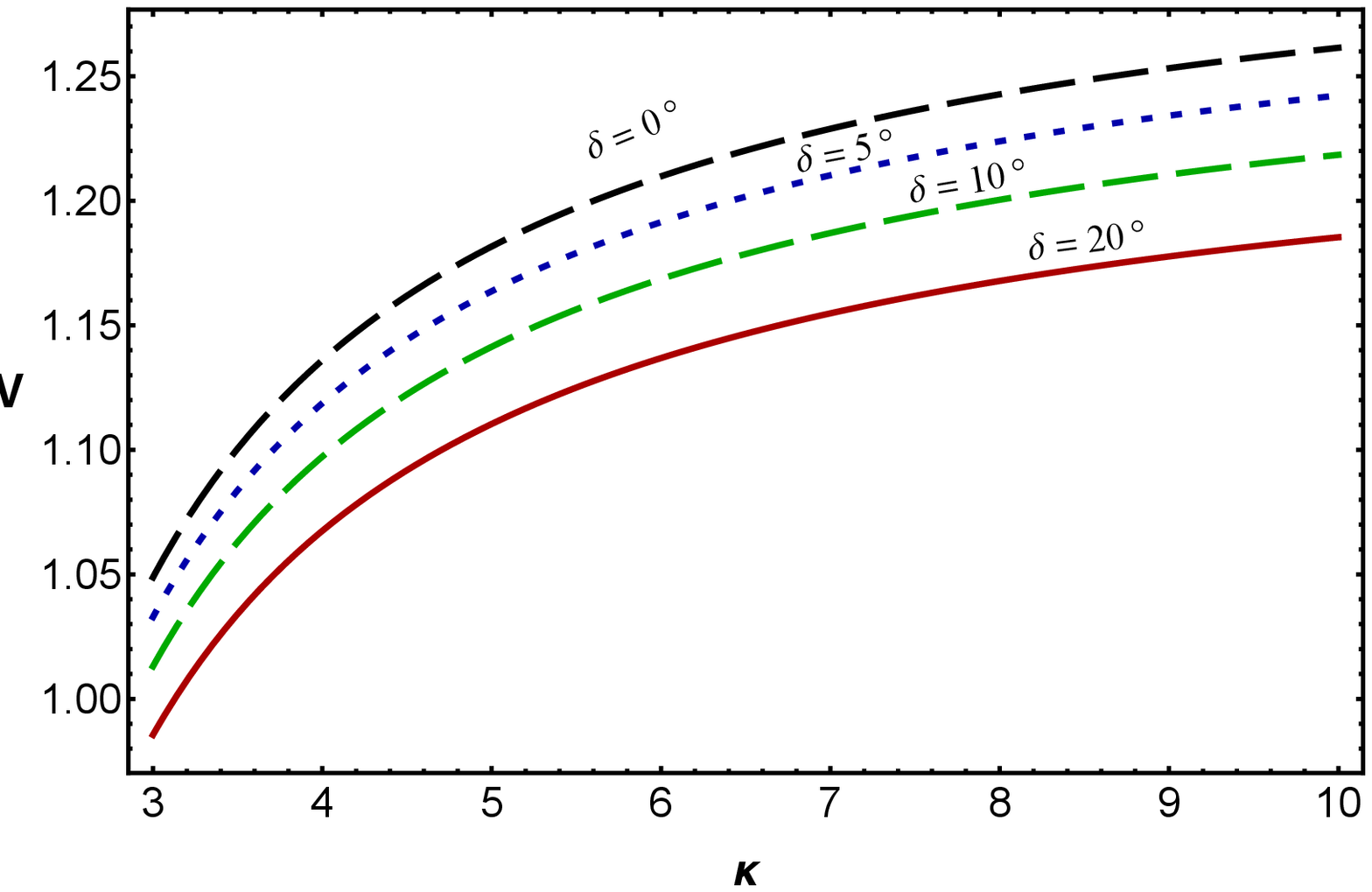}

\large{(a)}\vspace{0.6cm}

\includegraphics[width=6.7cm]{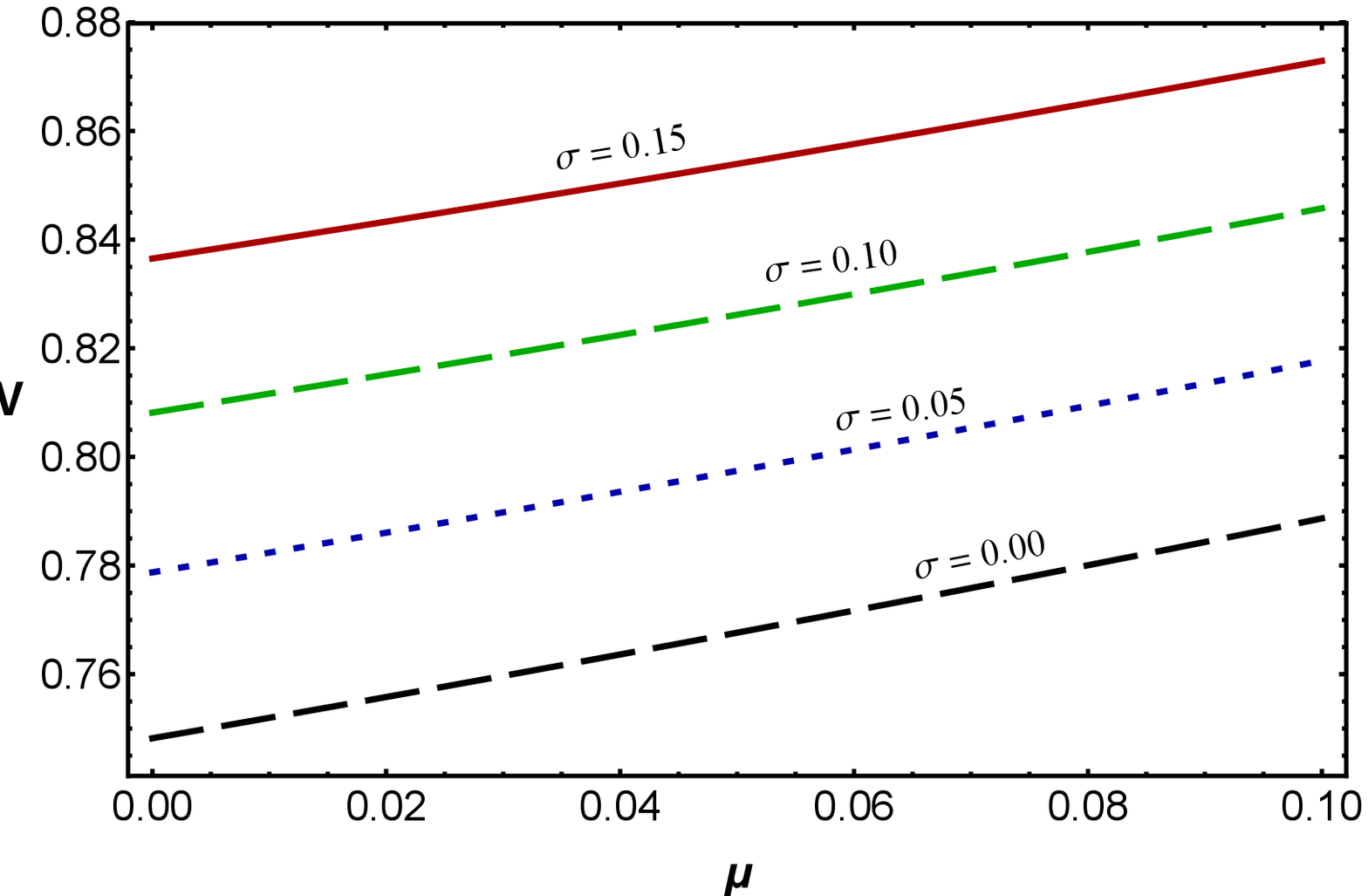}

\large{(b)}
\end{center}
\caption{(a) Phase speed $V$ versus the superthermality index $\kappa$ for different values of obliquity angle $\delta$,
where ion-to-electron temperature $\sigma=0.1$ and dust-to-ion number density $\mu=0.4$, and
(b) $V$ versus $\mu$ for different values of $\sigma$,
where $\kappa=3$ and $\delta=15\si{\degree}$.} \label{1fig:1}
\end{figure}

The variation of phase speed $V$ versus the superthermality index $\kappa$ for different values of obliquity angle $\delta$ is depicted
in Figure \ref{1fig:1}a. On the other hand, Figure \ref{1fig:1}b  shows the variation of $V$ versus the dust-to-ion number density
$\mu$ for different values ion-to-electron temperature $\sigma$.  We see in  Figure \ref{1fig:1}a that the phase speed of obliquely
propagating DAWs is higher in Maxwellian plasmas in comparison to that in superthermal plasmas, i.e., the phase speed in seen to lower
for smaller values of $\kappa$ (stronger superthermality) and it becomes higher for larger values  of $\kappa$
 (moderately nonthermal or Maxwellian). Figure \ref{1fig:1}a also suggests that when the DIAWs propagate through the plasma medium parallel (i.e.,
$\delta=0\si{\degree}$) to the magnetic field, they achieve the higher phase speed. However, when the waves advance obliquely (i.e., $\delta\neq0$),
the phase speed is predicted to decrease with the increase in obliquity angle. Figure \ref{1fig:1}b displays influence of the dust-to-ion number density $\mu$
and the ion-to-electron temperature $\sigma$ on the linear properties (i.e., the phase speed) of obliquely propagating DAWs. we found that the higher values
of $\mu$ leads to the formation/propagation of DAWs with higher phase speed in the considered plasma medium, while the increase (decrease) in ion (superthermal
electron) temperature may lead to propagate the DAWs with higher phase speed, as depicted in Figure \ref{1fig:1}b.

Now we derive the $x$ and $y$-components in terms of electric potential $\phi_1$ from the momentum equation by taking the same
coefficients as $z$-component, and they are given as
\begin{eqnarray}
\begin{rcases}
u_{1x}=-\bigg(\frac{l_y}{\Omega_{ci}}+\frac{\sigma l_yl_z^2}{\Omega_{ci}(V^2-\sigma l_z^2)}\bigg)\frac{\partial\phi_1}{\partial\xi},\\
u_{1y}=\bigg(\frac{l_x}{\Omega_{ci}}+\frac{\sigma l_xl_z^2}{\Omega_{ci}(V^2-\sigma l_z^2)}\bigg)\frac{\partial\phi_1}{\partial\xi}.
\end{rcases}
\label{1:eq14}\
\end{eqnarray}

We now separate the next order of $\epsilon$ from our model equations \eqref{1:eq6}-\eqref{1:eq8}, combine the resultant equations and
eliminate $n_2$, $u_{2x,y,z}$ and $\phi_2$ from the resultant equations (algebraic details are omitted here), and finally we consider
$\phi_1=\psi$ and get the time evolution equation in the form of damped KdV (dKdV) equation as
\begin{eqnarray}
&&\hspace*{-1.3cm}\frac{\partial\psi}{\partial\tau}+\alpha\psi\frac{\partial\psi}{\partial\xi}
+\beta\frac{\partial^3\psi}{\partial\xi^3}+\nu\psi=0,
\label{1:eq15}\
\end{eqnarray}
where the nonlinear term $\alpha$,  dispersion term $\beta$, and dissipation (or damped) term $\nu$ are given,
respectively, as follows
\begin{eqnarray}
&&\hspace*{-1.3cm}\alpha=V\bigg{(}p+\frac{p}{2(1+p\sigma)}-\frac{q}{p(1+p\sigma)}\bigg{)},
\label{1:eq16}\\
&&\hspace*{-1.3cm}\beta=\frac{V}{2p}\bigg{(}\frac{1}{(1+p\sigma)}+\frac{(1-l_z^2)(1+p\sigma)}{\Omega_{ci}}\bigg{)},
\label{1:eq17}\\
&&\hspace*{-1.3cm}\nu=\frac{\nu^{\prime}}{2}.
\label{1:eq18}\
\end{eqnarray}

It is well known that the mutual balance between the nonlinear term $\alpha$ and the dispersion term
$\beta$ form the solitary waves, where $\alpha$ determines the steepness/sharpness of the solitary excitations and
$\beta$ measures the broadening of the solitary waves. On the other hand, the dissipation/damping term $\nu$
measures the decay of the solitary wave over time while propagating. This is why it is essential
to thoroughly study the variations of these coefficients with the parameters on which they depend. From Eqs.
\eqref{1:eq16} and \eqref{1:eq17}, it is seen that $\alpha$ and $\beta$ both depend on some common plasma
parameters (such as the superthermality index $\kappa$, obliquity angle $\delta$, dust-to-ion number density
$\mu$, and ion-to-electron temperature $\sigma$, etc.). We see in equation \eqref{1:eq17} that the external
magnetic field $B_0$ has influence only on the dispersion term
$\beta$ via $\Omega_{ci}$ and the influence of $B_0$ diminishes as $l_z=\rm{cos}\theta=0^{\circ}$, i.e., for parallel propagation.

\begin{figure}[!h]
\begin{center}
\includegraphics[width=6.7cm]{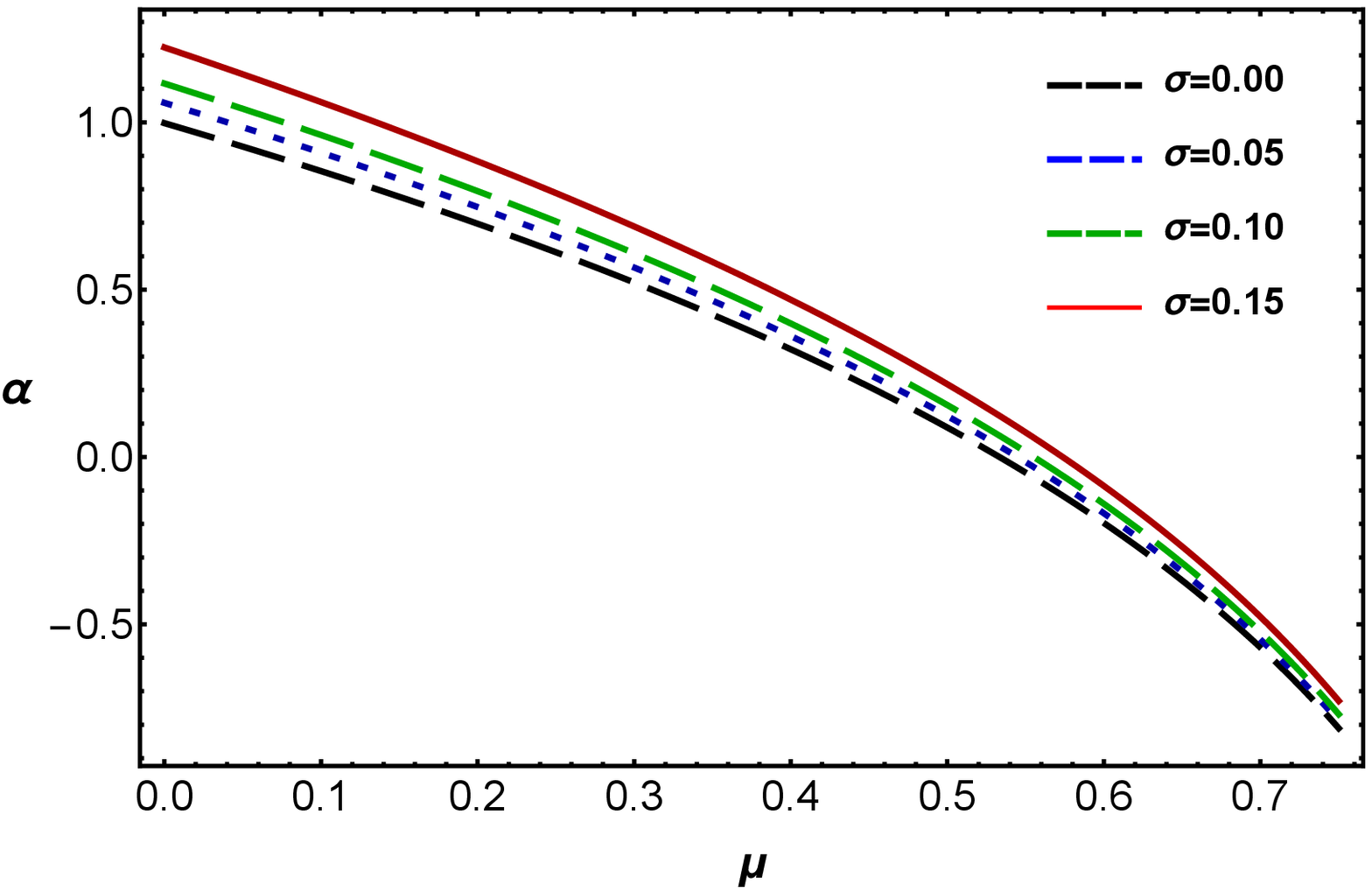}

\large{(a)}\vspace{0.6cm}

\includegraphics[width=6.7cm]{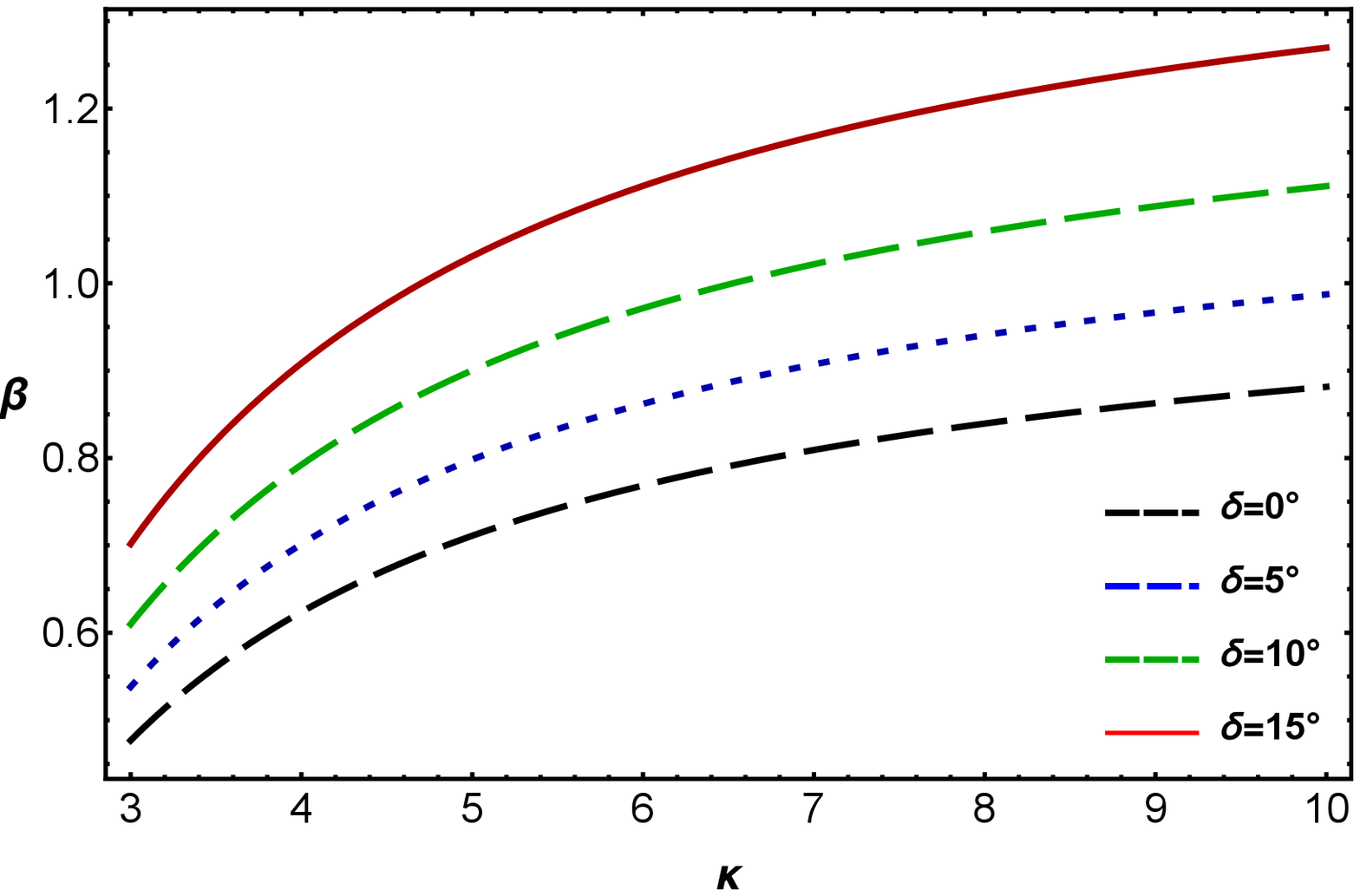}

\large{(b)}
\end{center}
\caption{The variation of (a) the nonlinearity coefficient $\alpha$ with $\mu$ for different values of
$\sigma$ for $\delta=?$ and $\kappa=?$, and (b) the dispersion coefficient $\beta$ with $\kappa$ for different
$\delta$, where $\mu=0.4$, $\sigma=0.10$. For both panel, we choose $\Omega_{ci}=0.5$.}
\label{1fig:2}\
\end{figure}

We now see how different plasma configuration parameters influence the nonlinear term $\alpha$ and the dispersion term  $\beta$. We now plot $\alpha$ against  $\kappa$ for different values of $\delta$ in Figure \ref{1fig:2}a to trace the influence of nonthermality and obliqueness while Figure \ref{1fig:2}b shows the variation of $\beta$ versus $\kappa$ for different values of obliquity angle via $\delta$.

\begin{figure}[!h]
\begin{center}
\includegraphics[width=6.7cm]{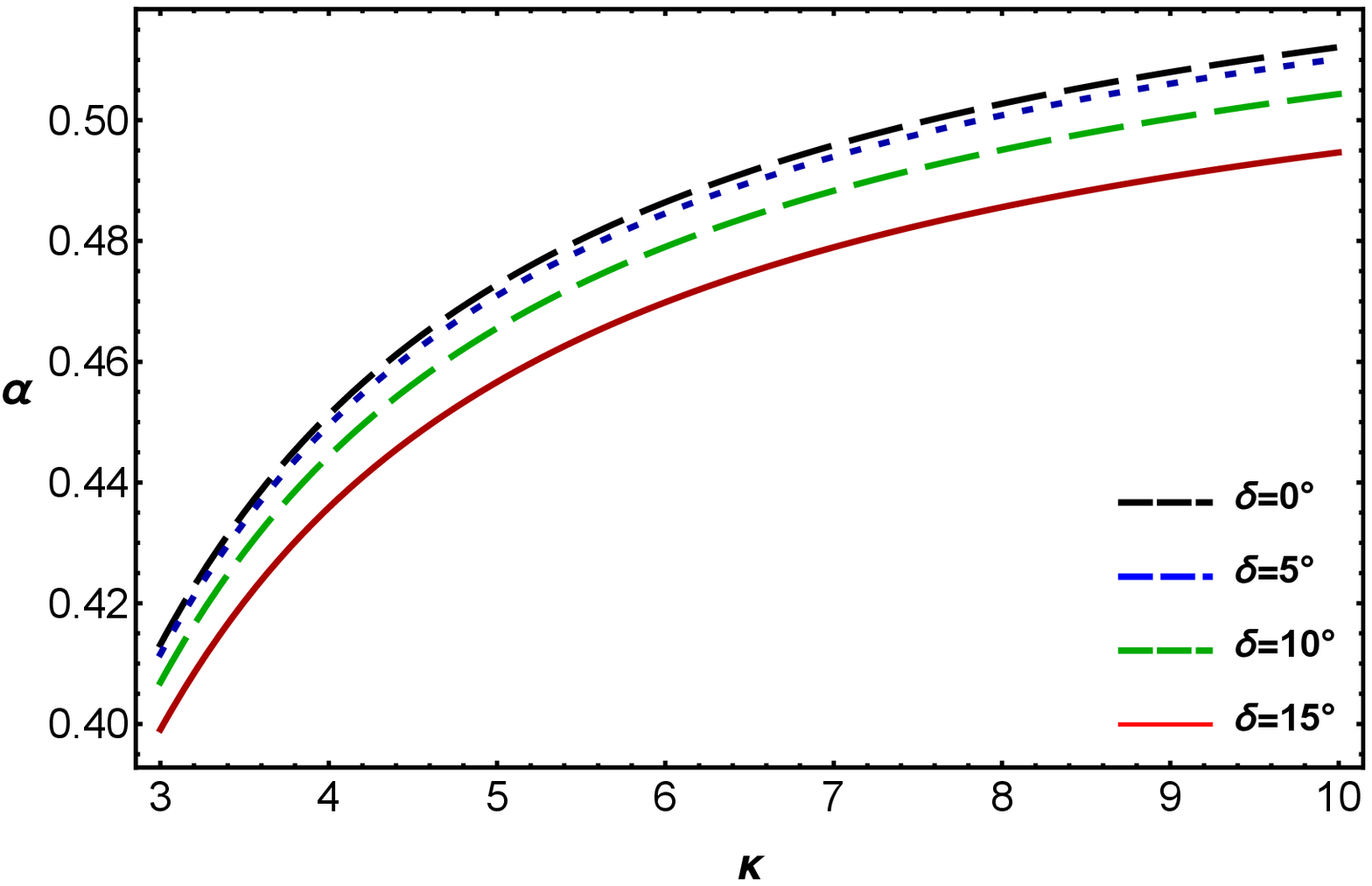}

\large{(a)}\vspace{0.6cm}

\includegraphics[width=6.7cm]{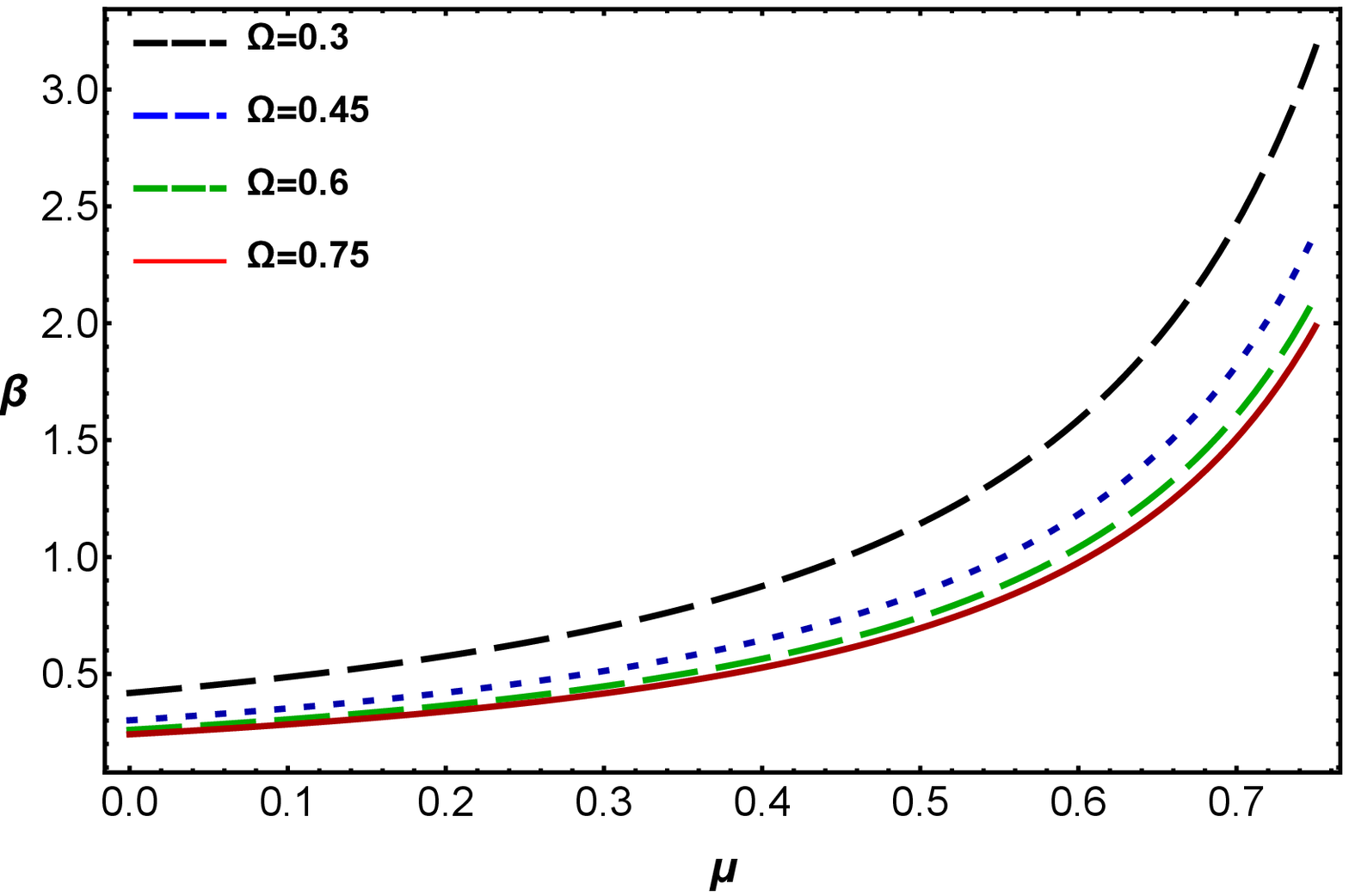}

\large{(b)}
\end{center}
\caption{The variation of (a) the nonlinearity coefficient $\alpha$ versus $\kappa$ for different values of $\delta$
with $\sigma=0.10$ and $\mu=0.4$, and (b) the dispersion coefficient $\beta$ versus $\mu$ for different
values of $\Omega_{ci}$ with $\kappa=3$, $\delta=15\si{\degree}$, and $\sigma=0.10$.}
\label{1fig:3}\
\end{figure}

 From these figures, it is clearly seen that both $\alpha$ and $\beta$ attain comparatively higher value in a thermally distributed (Maxwellian) plasma than in a superthermally distributed plasma, which suggests that the soliton formed in a Maxwellian plasma will be taller and broader than that in a superthermal plasma medium. Nevertheless, the coefficients show opposite characteristics for the variation of obliquity angle, $\delta$ i.e., nonlinearity decreases with obliquity angle while the dispersion increases as the value of the obliquity angle are increased. In Figure \ref{1fig:3}a, we depict the variation of $\alpha$ against $\mu$ for different values of ion and/or electron temperature via $\sigma$, and $\beta$ versus $\mu$ for different values of external magnetic field $B_0$ (via $\Omega_{ci}$) in Figure \ref{1fig:3}b. We see in Figure \ref{1fig:3}a that the value of $\alpha$ gets smaller for the higher value of $\mu$, i.e., the nonlinearity decreases with increasing dust number density compared to that of ion, and after a particular critical value, it downfalls.
 So our plasma model is valid for both dust ion-acoustic soliton of positive and negative potential. This figure also indicates that $\alpha$ increases with the increase of $\sigma$, which means that if the ion temperature increases with keeping the electron temperature fixed, the nonlinearity also rises for our considered medium. So the solitary structure will become less tall for the more significant value of ion to electron temperature ratio $\sigma$. Though the expression of $\beta$ contains $\sigma$, in reality, $\beta$ does not vary significantly with $\sigma$ while $\alpha$ does. So we have plotted $\beta$ against $\mu$ for various values of $\Omega_{ci}$ instead of $\sigma$ in Figure \ref{1fig:3}b, which shows $\beta$ increases with $\mu$, in contrast, decreases with $\Omega_{ci}$, i.e., the dispersion coefficient is seen to be increased with the ratio of dust to ion number density unlike the nonlinearity coefficient $\alpha$ and with the increase of the magnetic field $\textbf{B}_0$ the dispersive term gets decreased while the nonlinearity is not affected at all.

\begin{figure}[t!]
\centering
\includegraphics[width=70mm]{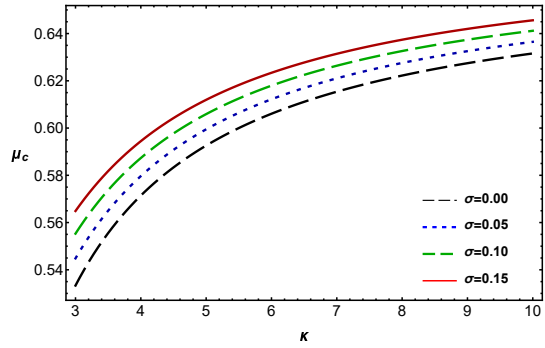}
\caption{Showing the variation of critical values of dust-to-ion number density $\mu_c$ with $\kappa$ for different $\sigma$ and for $\delta=15\si{\degree}$.}
\label{1Fig:4}\
\end{figure}

\begin{center}
\begin{table}[!h]
\caption{Critical values of dust-to-ion number density $\mu_c$ with and without
ion temperature effect, and for different values of $\kappa$}\vspace{0.1cm}
\begin{tabular}{   l   l   l   p{2cm}  }
    \hline\hline
     $\kappa$ \ \ \ \  \ \  \ \  & $\mu_c\,(\sigma\rightarrow0)$ \ \  \ \ \ & $\mu_c\,(\sigma\rightarrow0.15)$ \\ \hline
    $3$  & $0.533$ \ \ \  & $0.565$  \\ 
    $6$  & $0.606$ \ \  \ & $0.623$ \\ 
    $10$  & $0.632$ \ \ \  & $0.646$ \\ 
    $50$ & $0.660$ \ \ \ & $0.671$  \\
    $100$ & $0.663$ \  \ \ & $0.674$  \\
    \hline\hline
    \end{tabular}
\label{1Tab:1}
\end{table}
\end{center}

It is clear in equation \eqref{1:eq17} that the dispersion term $\beta$ is always positive for any given values of
plasma superthermality index $\kappa$, as expected; and the polarity of the DIASWs depends only on the
sign of the nonlinear term $\alpha$.  The dust-to-ion number density threshold (i.e., the critical number density
of dust-to-ion), we define as $\mu_c$, can be obtained by solving $\alpha[l_z,\sigma,\mu,\kappa]=0$ for $\mu$, has the form

\begin{eqnarray}
&&\hspace*{-1.3cm}\mu_c=\frac{1}{16\kappa^2\sigma-16\kappa\sigma+4\sigma}
\times\bigg[(16\sigma+12)\kappa^2-(16\sigma+24)\kappa+(4\sigma+9)\pm(2\kappa-1)
\times\sqrt{(2\kappa-3)\big\{(16\sigma+18)\kappa+(8\sigma-27)\big\}}\bigg].
\label{1:eq19}
\end{eqnarray}

It is worth to note that the nonlinear term $\alpha>0$ for $\mu>\mu_c$ and $\alpha<0$ for $\mu<\mu_c$.
We see in equation (\ref{1:eq19}) that the critical number threshold $\mu_c$, depending on which the considered dusty plasma
medium may form the positive or negative potential solitary excitations, is an explicit function of the superthermality parameter $\kappa$
and the ion-to-electron temperature $\sigma$. We depict the variation of critical values of dust-to-ion number density $\mu_c$ with
the superthermality parameter $\kappa$ for different values of ion or electron temperature via $\sigma\,(=T_i/z_iT_e)$. We found that
the critical number density is seen to increase with the increase (decrease) in ion (electron) temperature for the fixed values of $z_i$,
while $\mu_c$ is smaller (in magnitude) in superthermal (lower $\kappa$ values) plasmas in comparison to that in moderately
non-thermal or Maxwellian (higher $\kappa$ values) plasmas, as shown in Figure \ref{1Fig:4}.

\textbf{However, at $\mu=\mu_c$, the amplitude of the solitary wave becomes infinite, which compels us to consider a new stretching to study the solitary waves for this particular condition. Thus, we have introduced a new set of stretching coordinates as}
\begin{eqnarray}
\begin{rcases}
\xi=\epsilon^{1/2}(l_xx+l_yy+l_zz-Vt)),\\ \tau=\epsilon^{3/2}t,\\ \nu_i=\epsilon^3\nu^\prime,
\end{rcases}
\label{1:eq19a}\
\end{eqnarray}
\textbf{to derive the damped modified KdV (dmKdV) equation with the help of the following set of variable expansions:}

\begin{eqnarray}
\begin{pmatrix}n\\u_{x,y}\\u_z\\ \phi\end{pmatrix}=\begin{pmatrix}1\\0\\0\\0\end{pmatrix}+
   \epsilon\begin{pmatrix}n_1\\ \epsilon u_{1x,y}\\u_{1z}\\ \phi_1\end{pmatrix}
+\epsilon^2\begin{pmatrix}n_2\\ \epsilon u_{2x,y}\\u_{2z}\\ \phi_2\end{pmatrix}
+\epsilon^3\begin{pmatrix}n_3\\ \epsilon u_{3x,y}\\u_{3z}\\ \phi_3\end{pmatrix}+\cdots
\label{1:eq19b}\
\end{eqnarray}
\textbf{Using \eqref{1:eq19a}-\eqref{1:eq19b} and then performing the similar method as before, the following dmKdV equation can be derived efficiently.
\begin{eqnarray}
&&\hspace*{-1.3cm}\frac{\partial\phi_1}{\partial\tau}+A\phi_1^2\frac{\partial\phi_1}{\partial\xi}
+B\frac{\partial^3\phi_1}{\partial\xi^3}+C\frac{\partial\phi_1\phi_2}{\partial\xi}+D\phi_1=0,
\label{1:eq19c}\
\end{eqnarray}
where
\begin{eqnarray}
&&\hspace*{-1.3cm}A=V\bigg{(}3q-6p^2+\frac{pq-3r}{2p(1+p\sigma)}\bigg{)},
\label{1:eq19d}
\end{eqnarray}
It is absolute from the coefficients that the dispersion coefficient $B$ and the collision coefficient $D$ remain exactly the same as determined before. At threshold dust composition (i.e., $\mu=\mu_c$), $\alpha=C=0$. Now if we consider, $A=\alpha^\prime$ and $\phi_1=\psi^\prime$, equation \eqref{1:eq19c} results as:}
\begin{eqnarray}
&&\hspace*{-1.3cm}\frac{\partial\psi^\prime}{\partial\tau}+\alpha^\prime\psi^{^\prime2}\frac{\partial\psi^\prime}{\partial\xi}
+\beta\frac{\partial^3\psi^\prime}{\partial\xi^3}+\nu\psi^\prime=0,
\label{1:eq19e}\
\end{eqnarray}

\begin{figure}[t!]
\centering
\includegraphics[width=70mm]{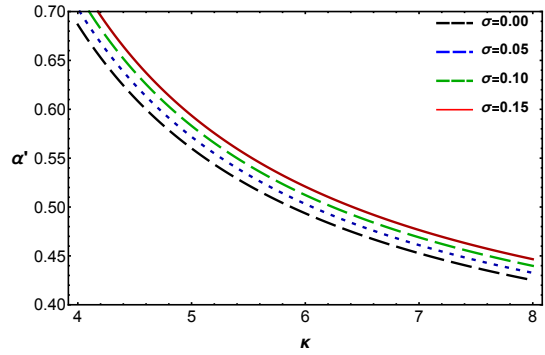}
\caption{The variation of the higher order nonlinearity coefficient (a) $\alpha^\prime$ versus $\kappa$ for different values of $\sigma$
with $\delta=10\si{\degree}$ and $\mu=\mu_c$.}
\label{1Fig:4a}\
\end{figure}
\textbf{It is explicit from Figure \ref{1Fig:4a} that the nonlinearity constant $\alpha^\prime$ is always positive for different plasma parameters at $\mu=\mu_c$. Thus we can predict that at critical dust composition, only compressive solitary waves can exist and propagate in our considered plasma medium.}

\subsection{Analytical solution}
\label{1sec:ANALYTICAL SOLUTION}
If we consider $\nu=0$ (i.e., there is no collision between ions and neutral particles), equation \eqref{1:eq15} reduces
to the standard form of KdV equation as
\begin{eqnarray}
&&\hspace*{-1.3cm}\frac{\partial\psi}{\partial\tau}+\alpha\psi\frac{\partial\psi}{\partial\xi}
+\beta\frac{\partial^3\psi}{\partial\xi^3}=0,
\label{1:eq20}\
\end{eqnarray}
upon which the integration can be performed completely and it follows an infinite set of conservation laws.
If $E$ is assumed as the soliton's energy, the energy conservation is then expressed as
\begin{eqnarray}
&&\hspace*{-1.3cm}\frac{\partial E}{\partial\tau}=0,
\label{1:eq21}\
\end{eqnarray}
where
\begin{eqnarray}
&&\hspace*{-1.3cm}E=\int_{-\infty}^{\infty}\psi^2(\xi,\tau)d\xi
\nonumber\
\end{eqnarray}
and the solution of equation \eqref{1:eq20} is
\begin{eqnarray}
&&\hspace*{-1.3cm}\psi(\xi,\tau)=\Psi\textmd{sech}^2\bigg[\sqrt{\frac{\alpha\Psi}{12\beta}}\big(\xi-\frac {\alpha\Psi}{3}\tau\big)\bigg].
\label{1:eq22}\
\end{eqnarray}
\textbf{By substituting equation \eqref{1:eq22} in the energy expression and then performing the integration gives}
\begin{eqnarray}
&&\hspace*{-1.3cm}E=\frac{4}{3}\times\Psi^2\times\sqrt{\frac{4\alpha}{U_0}}.
\label{1:eq22a}
\end{eqnarray}
The conditions for the formation of localized structures
\begin{eqnarray}
\begin{rcases}
\psi\rightarrow 0,\\
\frac{\partial\psi}{\partial\xi}\rightarrow 0,\\
\frac{\partial^2\psi}{\partial\xi^2}\rightarrow 0
\end{rcases}
\label{1:eq23}\
\end{eqnarray}
are fulfilled as $\xi\rightarrow\pm\infty$. In Eq. \eqref{1:eq21}, $\Psi(=\frac{3U_0}{\alpha})$ is the soliton amplitude,
$L(=\sqrt{\frac{4\alpha}{U_0}}=\sqrt{\frac{12\beta}{\alpha\Psi}})$ is the soliton width, and $U_0(=\frac{\alpha\Psi}{3})$
is the soliton speed. That is, the KdV equation (\ref{1:eq20}) is fully integrable and one can find the exact solution of
KdV equation when there is no collision or damping is present in the plasma medium (i.e., $\nu^\prime=0$).
In the presence of collision or damping/dissipation (i.e., $\nu^\prime\neq0$), it is not possible to solve the dKdV equation (\ref{1:eq15})
analytically for an exact solution and one has to consider an approximate solution for the weak dissipation due to ion-neutral
collisional effect (i.e., for the assumption $\nu^\prime\ll0$). We note that equation \eqref{1:eq15} is not a completely integrable Hamiltonian
system, i.e.,  the soliton energy $E$ will definitely not be conserved, and
\begin{eqnarray}
&&\hspace*{-1.3cm}\frac{\partial E}{\partial\tau}=-\nu^\prime E,
\nonumber\\
&&\hspace*{-1.3cm}\Rightarrow E(\tau)=E(0)\:\textmd{e}^{-\nu^\prime\tau}.
\nonumber\
\end{eqnarray}
The time dependent form of the soliton energy, amplitude, speed, and width \cite{Sahu2017,Sultana2018}) are, respectively, expressed as
\begin{eqnarray}
&&\hspace*{-1.3cm}\psi_m(\tau)=\psi(0)\:\textmd{e}^{-\frac{2\nu^\prime}{3}\tau},
\nonumber\\
&&\hspace*{-1.3cm}U_0(\tau)=U_0(0)\:\textmd{e}^{-\frac{2\nu^\prime}{3}\tau},
\nonumber\\
&&\hspace*{-1.3cm}L(\tau)=\sqrt{\frac{12\beta}{\alpha\,\psi(0)}}\:\textmd{e}^{\frac{\nu^\prime}{3}\tau},
\nonumber\
\end{eqnarray}
and the approximate analytical process gives the solution of equation \eqref{1:eq15}  as \cite{Karpman1977,Herman1990,Sahu2017}
\begin{eqnarray}
&&\hspace*{-1.3cm}\psi_{\nu}(\xi,\tau)=\psi_m(\tau)\textmd{sech}^2\bigg[\sqrt{\frac{\alpha\,\psi_m(\tau)}{12\beta}}\big(\xi-
\frac{\alpha}{3}\int_{0}^{\tau}\psi_m(\tau)d\tau\big)\bigg],
\nonumber\\
&&\hspace*{-1.3cm}\Rightarrow\psi_{\nu}(\xi,\tau)=\psi(0)\textmd{e}^{\frac{-2\nu^\prime}{3}\tau}\textmd{sech}^2\bigg[\sqrt{\frac{U_0(\tau)}{4\beta}}
\bigg\{\xi-\frac{\alpha\,\psi(0)}{2\nu^\prime}\big(1-\textmd{e}^{\frac{-2\nu^\prime}{3}\tau}\big)\bigg\}\bigg],
\label{1:eq24}\
\end{eqnarray}
where $\psi(0)$ is the soliton amplitude at time $\tau=0$, which would remain the same with time unless the collisional
parameter is present in the system. As the time evaluates, the soliton gets damped due to the term $\nu^\prime(=2\nu)$ and eventually
diminishes over time while propagating.

\textbf{Similar to the dKdV equation, mdKdV also does not have any exact analytical solution. However, in the absence of collision (i.e., $\nu=0$), the modified KdV equation has a straightforward analytical solution \cite{Wadati1975} as
\begin{eqnarray}
&&\hspace*{-1.3cm}\psi^\prime(\xi,\tau)=\Psi^\prime\textmd{sech}\bigg(\frac{\xi-U_0\tau}{L^\prime}\bigg),
\label{1:eq24a}\
\end{eqnarray}
where amplitude and width are defined as $\Psi^\prime=\sqrt{6U_0/\alpha^\prime}$ and $L^\prime=\sqrt{\beta/U_0}$.
As there is no exact analytical solution for dmKdV, we have followed the semi-analytical approach of Kashkari et al. \cite{Kashkari2021} and developed the the following solution
\begin{eqnarray}
&&\hspace*{-1.3cm}\psi_\nu^\prime(\xi,\tau)=\Psi_m^\prime\textmd{sech}\bigg(\frac{\Theta^\prime(\tau)}{L^\prime(\tau)}\bigg),
\label{1:eq24b}\
\end{eqnarray}
where
\begin{eqnarray}
&&\hspace*{-1.3cm}\Theta^\prime(\tau)=\xi-UH(\tau),
\nonumber\\
&&\hspace*{-1.3cm}H(\tau)=\frac{2}{3\nu}e^{-\nu\tau}(e^{\frac{3\nu}{2}\tau}-1),
\nonumber\\
&&\hspace*{-1.3cm}\psi^\prime_m(\tau)=\psi^\prime_m(0)e^{-\nu\tau},
\nonumber\\
&&\hspace*{-1.3cm}L^\prime(\tau)=L^\prime(0)e^{\frac{\nu}{2}\tau}.
\nonumber\
\end{eqnarray}
Here, it is worth noting that at the limiting condition $\nu\rightarrow0$, this solution in equation \eqref{1:eq24b} will reduce to the analytical solution \eqref{1:eq24a}.}

\subsection{Characteristics of damped DIASWs}\label{1sec:Numerical simulation}
We are now interested to examine the characteristics of obliquely propagating damped DIASWs in a magnetized collisional $\kappa-$nonthermal dusty
plasma medium, in order to trace the effect of different plasma compositional parameters
(especially, the effect of ion temperature via $\sigma$, the damping effect via $\nu$, the plasma nonthermality or superthermality
via $\kappa$, the effect of ion and dust density $\mu$, and the influence of obliqueness via $\delta$, etc.) on the dynamics of damped DIASWs.
An approximate solution of damped KdV equation \eqref{1:eq15} for weak damping ($\nu\ll1$) [given in equation
(\ref{1:eq24})] is used to analyse the nature of obliquely propagating damped DIASWs parametrically. It is already mentioned in
Sec. \ref{1sec:ANALYTICAL SOLUTION} that in the presence of ion-neutral collision (i.e., $\nu\neq0$), it is no longer a
completely integrable Hamiltonian system. We, therefore, consider the solitary wave solution of equation \eqref{1:eq15} in the absence of collision is
$\psi(\xi,0)=\Psi\textmd{sech}^2\bigg[\sqrt{\frac{\alpha\Psi}{12\beta}}\xi\bigg]$, to analyze the DIASWs in a collisionless plasma. However,  the approximate solution in Eq. \eqref{1:eq24} is used to examine the dynamics of obliquely propagating damped DIASWs in the considered plasma medium.

\begin{figure}[!h]
\begin{center}
\includegraphics[width=6.7cm]{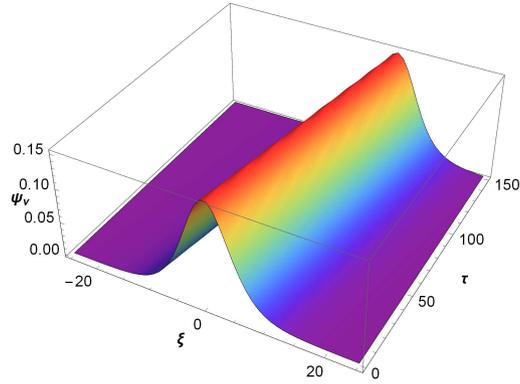}

\large{(a)}\vspace{0.6cm}

\includegraphics[width=6.7cm]{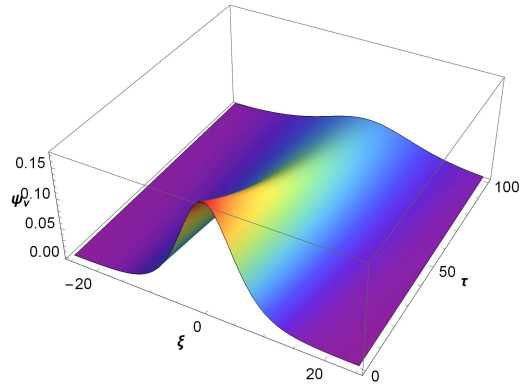}

\large{(b)}\vspace{0.6cm}

\includegraphics[width=6.7cm]{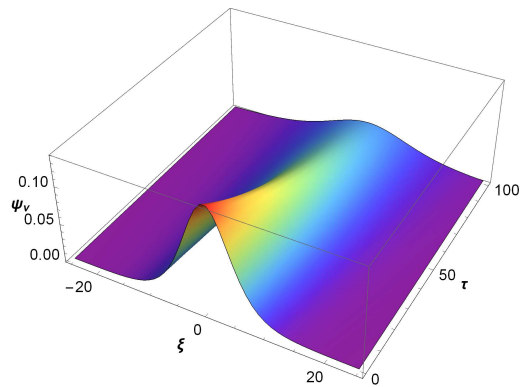}

\large{(c)}
\end{center}
\caption{Evolution of dust-ion-acoustic solitons [given in equation (\ref{1:eq24})] in a plasma with (a) $\nu=0$ and $\sigma=0$, (b) $\nu=0.01$ and $\sigma=0$, and
(c) $\nu=0.01$ and $\sigma=0.20$ . Other plasma parameters are fixed at $U_0=0.05$, $\mu= 0.1$, $\delta=10\si{\degree}$, $\Omega_{ci}=0.2$, $\sigma=0.10$, and $\kappa=3$.}
\label{1fig:5}\
\end{figure}

\begin{figure}[!h]
\begin{center}
\includegraphics[width=6.7cm]{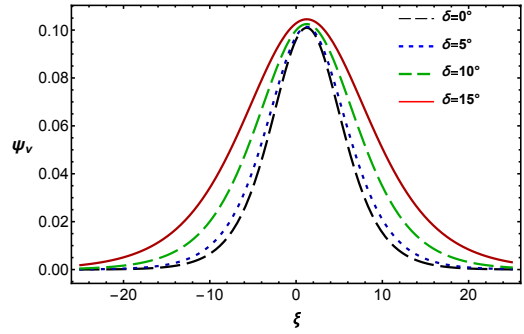}\end{center}
\caption{Dust-ion-acoustic solitary potential  $\psi_{\nu}$ [given in equation (\ref{1:eq24})] versus $\xi$ for different $\delta$ at $\tau=30$ in a collisional magnetized
nonthermal dusty plasma for $\mu=0.1$, $\Omega_{ci}=0.2$, $\sigma=0.10$, $U_0=0.05$,
$\kappa=3$, $\nu=0.01$.}
\label{1Fig:6}\
\end{figure}

\begin{figure}[!h]
\begin{center}
\includegraphics[width=6.7cm]{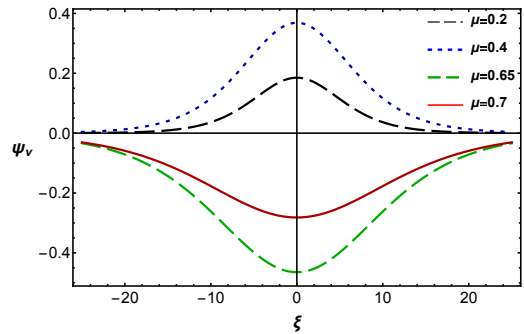}\end{center}
\caption{Positive potential DIASWs for $\mu<\mu_c$ and negative potential DISWs for $\mu>\mu_c$ at $\tau=0$ in the considered
plasma for $\delta=15\si{\degree}$, $\Omega_{ci}=0.2$, $\sigma=0.10$, $U_0=0.05$,
$\kappa=3$, $\nu = 0.01$.}
\label{1Fig:7}\
\end{figure}

In Figure \ref{1fig:5}a, we depict the evolution of DIASWs in the a collisionless dusty plasma (i.e., $\nu=0$) with $\kappa=3,\,\mu=0.1,\,\delta=10\si{\degree},\,
\Omega_{ci}=0.2,\,U_0=0.05$.
We found that the DIASW is seen to maintain it's stability over time while propagating, i.e., the amplitude and
the width of the pulse remain constant with time while advancing through a collisionless magnetized dusty plasma.  On the other hand, solitary wave
solution  (\ref{1:eq22}) is considered to propagate in a magnetized superthermal plasma in absence of ion temperature effect (i.e., $\sigma=0$) but in the presence of  collision effect ($\nu=0.01$), and for $U_0=0.05$,
$\mu=0.1$, $\delta=10\si{\degree}$, $\Omega_{ci}=0.2$, and $\kappa=3$,  to see the influence of dissipation due to collision only, as depicted in Figure \ref{1fig:5}b. It is expected and also clear from Figure  \ref{1fig:5}b that the amplitude (width) of the obliquely propagating DIASWs decreases (increases) with time and thus, the soliton property ($amplitude\times width^2=constant$) remains conserved. In Figure \ref{1fig:5}c, we examine the dissipation due to both the ion temperature and the collisional effect, where the other parametric values are same as in Figure \ref{1fig:5}b. The combined effect of these two parameters consequences in a more prominent dissipation of the soliton than that with considering the collision only.

We now see the effect of obliqueness $\delta$ in Figure \ref{1Fig:6} and dust-to-ion number density $\mu$ in Figure \ref{1Fig:7}.
Figure \ref{1Fig:6} shows the geometrical characteristics of the
solitary structures in space for different obliquity angle in such a plasma medium which has the following parameters $\mu=0.1$,
$\Omega_{ci}=0.2$, $\sigma=0.10$, $U_0=0.05$, $\kappa=3$, $\nu=0.01$ and $\tau=30$.
The plot suggests that both the amplitude and width of the pulse increase with the obliquity angle i.e., when the pulse propagates along the
external magnetic field $B_0$ ($\delta=0\si{\degree}$), amplitude and width have the smallest values and as the value of obliquity
angle increases, both the amplitude and width also increase which means the solitary structure becomes taller and wider
with the increase of obliquity angle. On the other hand, the influence of dust-to-ion number density on the propagation nature/characteristics
of obliquely propagating damped DIASWs is studied in the considered plasma for fixed plasma parameters $\delta=15\si{\degree}$, $\Omega_{ci}=0.2$,
$\sigma=0.10$, $U_0=0.05$, $\kappa=3$, $\nu=0.01$, and $\tau=0$. We have seen that our considered plasma system allows the propagation of
positive potential DIASWs for $\mu<\mu_c$ and negative potential  DIASWs for $\mu>\mu_c$, and both the amplitude and the width of the observed DIASWs
are seen to increase with the increase (decrease) of $\mu$ for the positive (negative) potential solitary excitation region, as depicted in Figure \ref{1Fig:7}.

\begin{figure}[!h]
\begin{center}
\includegraphics[width=6.7cm]{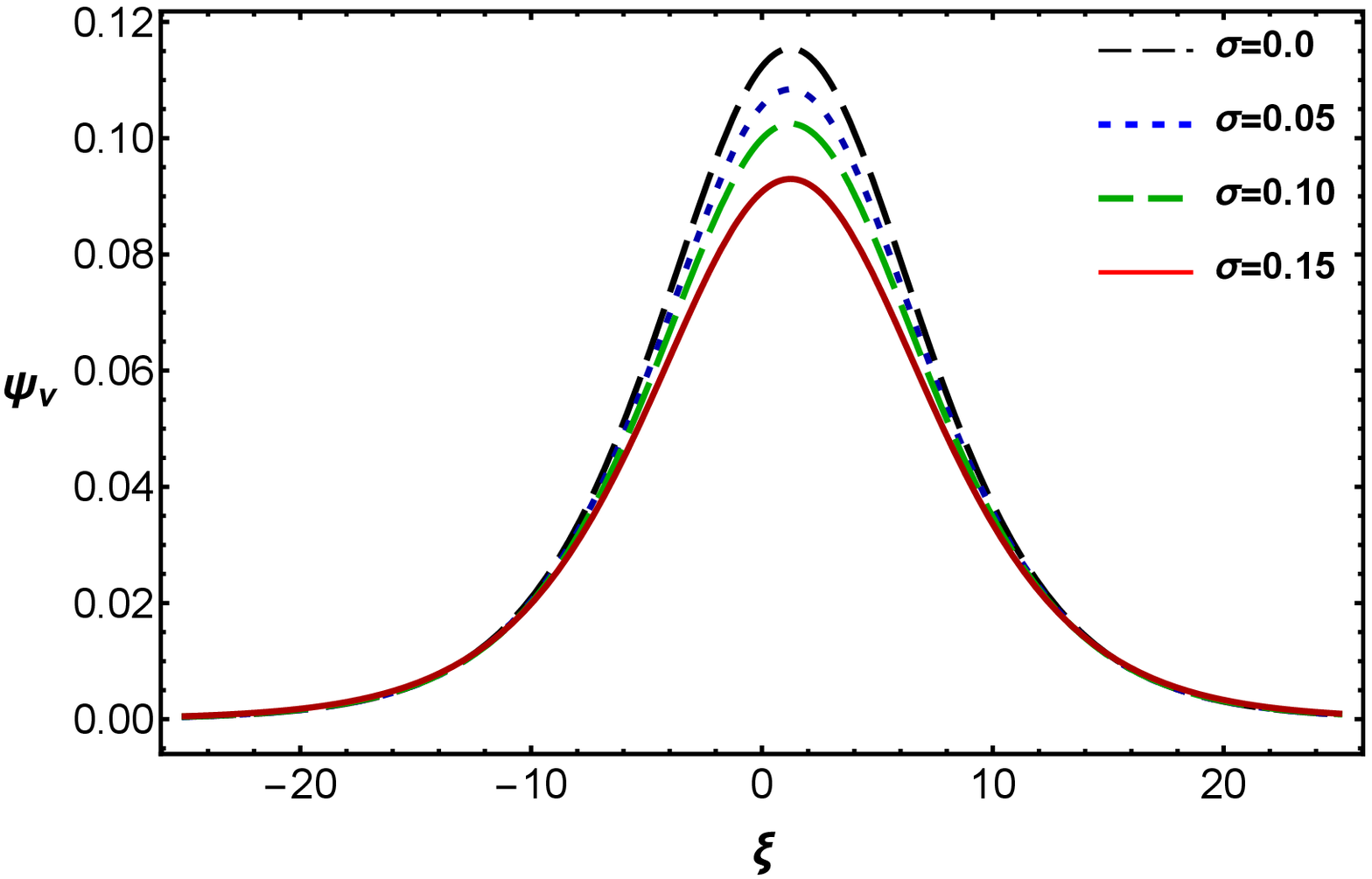}

\large{(a)}\vspace{0.6cm}

\includegraphics[width=6.7cm]{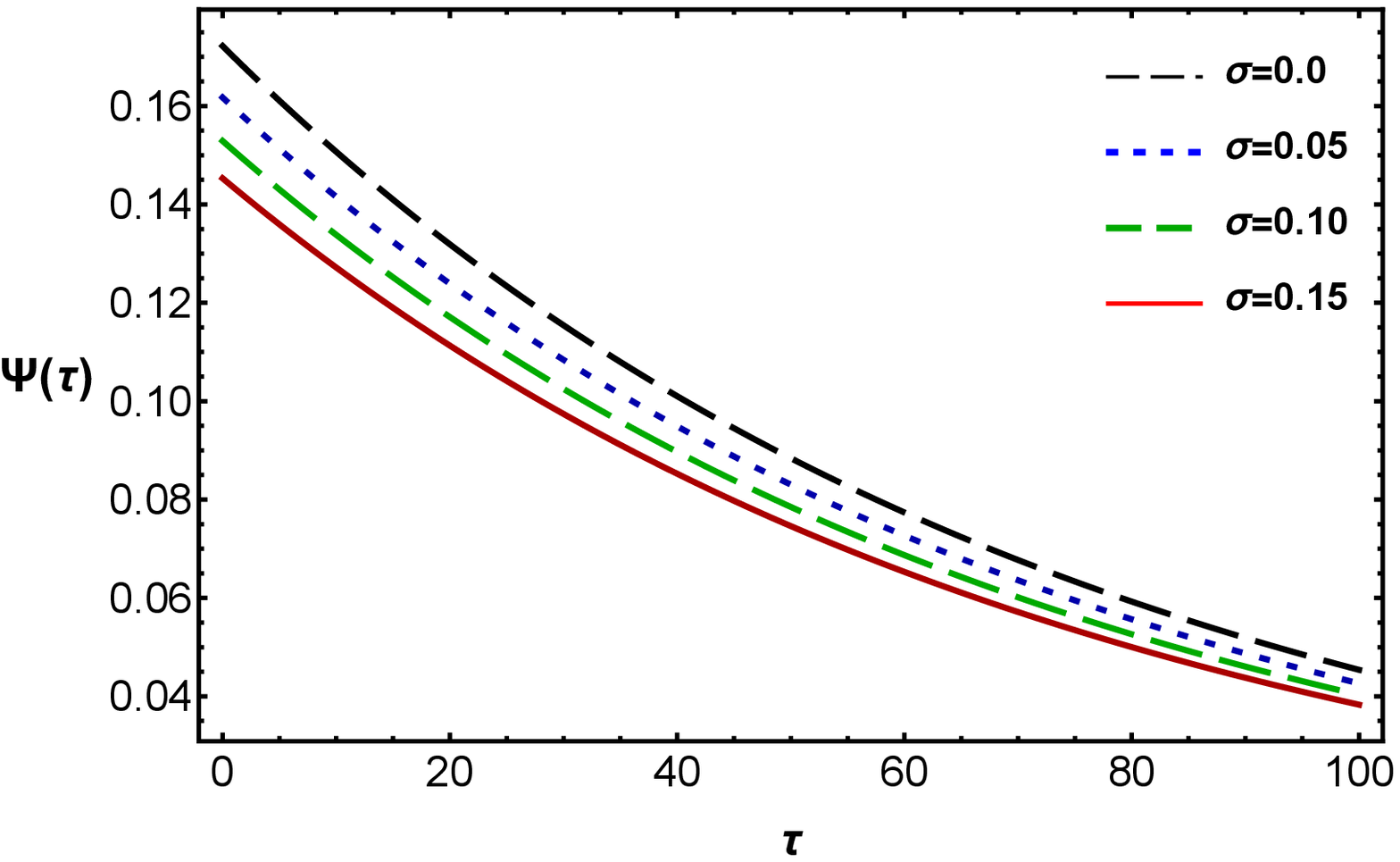}

\large{(b)}\vspace{0.6cm}

\includegraphics[width=6.7cm]{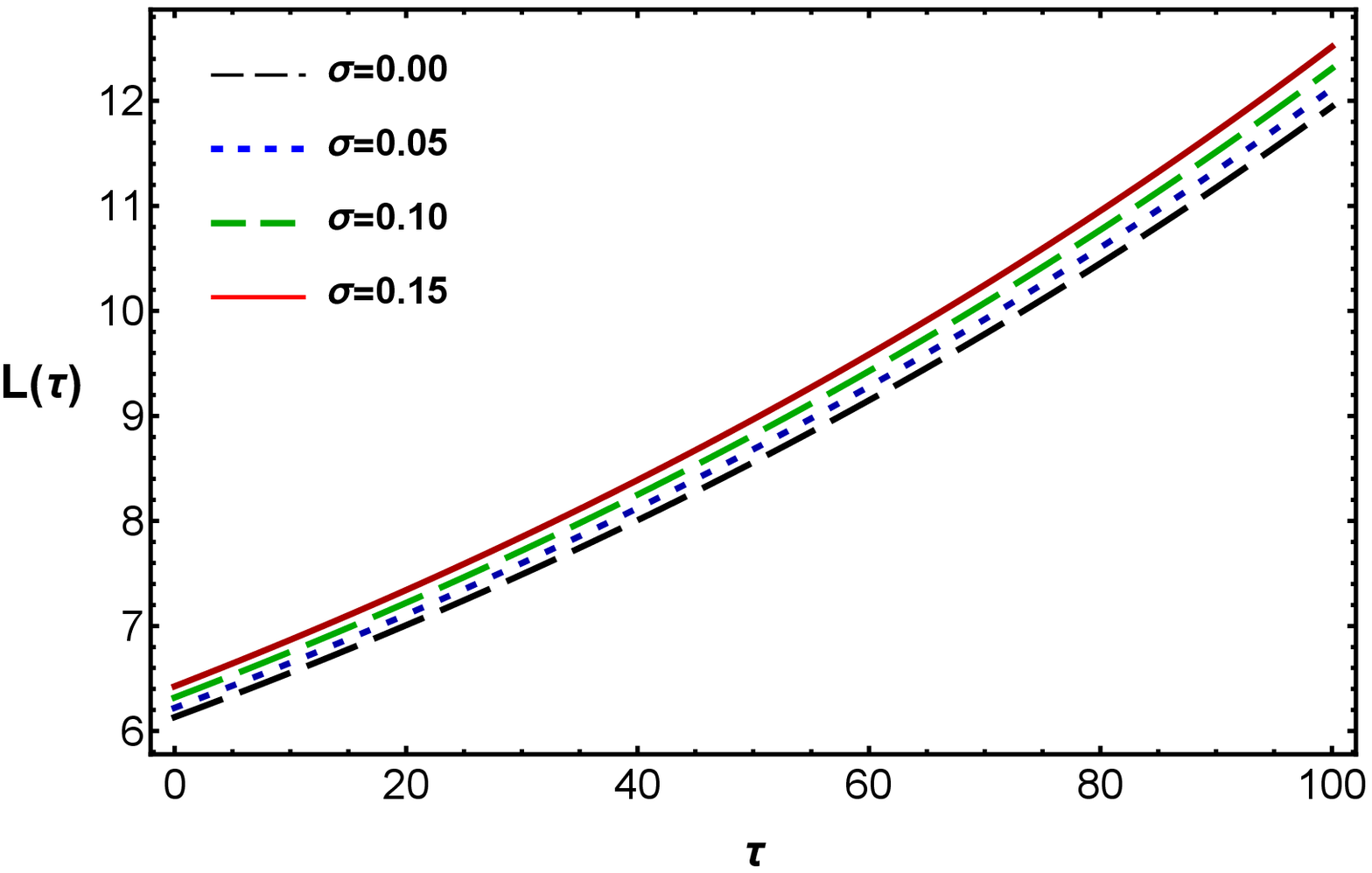}

\large{(c)}
\end{center}
\caption{Showing the variation of (a) $\psi_{\nu}$ versus $\xi$ at time $\tau=30$, (b) damped DIASWs amplitude $\psi_m(\tau)$ versus $\tau$, and
(c) damped DIASWs width $L(\tau)$ versus $\tau$, for different $\sigma$; where other plasma compositional parameters are fixed at $\delta=15\si{\degree}$,
$\Omega_{ci}=0.2$, $\mu=0.1$, $U_0=0.05$, $\kappa=3$, $\nu= 0.01$.}
\label{1fig:8}\
\end{figure}

\begin{figure}[!h]
\begin{center}
\includegraphics[width=6.7cm]{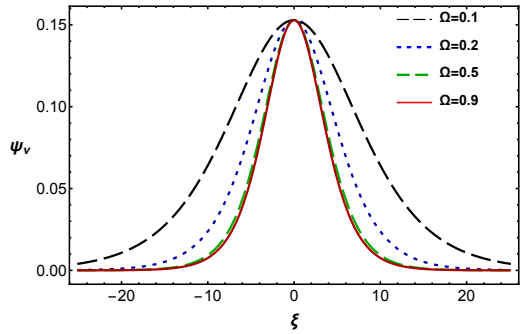}\end{center}
\caption{Effect of external magnetic field $B_0$ (via $\Omega_{ci}$) on damped DIASWs at $\tau=0$ in a magnetized collisional nonthermal
plasma for $\delta=15\si{\degree}$, $\sigma=0.1$, $\mu=0.1$, $U_0=0.05$,
$\kappa=3$, $\nu = 0.01$.}
\label{1Fig:9}\
\end{figure}

The influence of ion (nonthermal electron) temperature via $\sigma\,(=T_i/z_iT_e)$ on the dynamical properties of obliquely propagating dust-ion-acoustic solitary structures is analysed in Figure  \ref{1fig:8} in a collisional magnetized plasma medium in which $\delta=15\si{\degree}$, $\Omega_{ci}=0.2$, $\mu=0.1$, $U_0=0.05$, $\kappa=3$, $\nu=0.01$, and $\tau=0$. It is examined  that as the ion temperature effect increases on the
plasma medium, it causes the DIAW solitary waves to become smaller (in amplitude) and wider (in width), which is clearly depicted in Figures \ref{1fig:8}b and \ref{1fig:8}c. It is predicted from Figure \ref{1fig:8} that taller the solitary excitation the narrower it will be, as expected from (\ref{1:eq24}). We now see the effects of external magnetic field $B_0$ (via $\Omega_{ci}$)  in Figure \ref{1Fig:9} in a nonthermal collisional magnetized plasma with compositinal parameters $\delta=15\si{\degree}$, $\sigma=0.1$, $\mu=0.1$, $U_0=0.05$, $\kappa=3$, $\nu=0.01$, and $\tau=30$. It is found that the external magnetic field does not have any effect on the amplitude of DIASWs, but $B_0$ has a significant effect on the width of DIASWs, which is clearly depicted in Figure \ref{1Fig:9} and agrees with previous research in collisional and collisionless plasma contexts \cite{Singh2006}.

\begin{figure}[!h]
\begin{center}
\includegraphics[width=6.7cm]{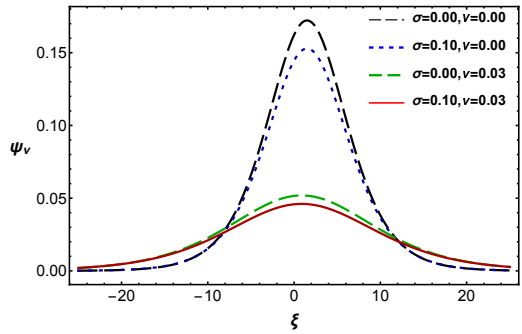}\end{center}
\caption{Showing the variation of solitary profile $\psi_{\nu}$ versus $\xi$ for different $\sigma$ and $\nu$, where $\mu=0.1$, $\Omega_{ci}=0.2$, $\delta=10\si{\degree}$, $U_0=0.05$,
$\kappa=3$, and $\tau=30.$}
\label{1Fig:10}\
\end{figure}

\begin{figure}[!h]
\begin{center}
\includegraphics[width=6.7cm]{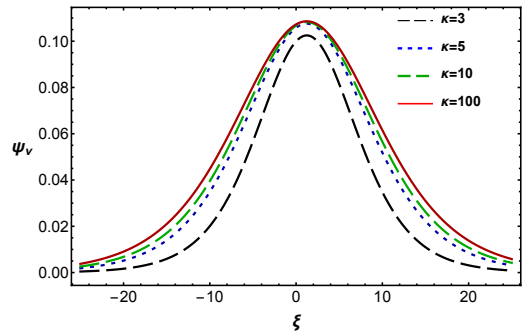}\end{center}
\caption{Variation of $\psi_{\nu}$ with $\xi$ for different $\kappa$, where $\delta=15\si{\degree}$, $\sigma=0.1$, $\mu=0.1$, $U_0=0.05$,
$\Omega_{ci}=0.2$, $\nu = 0.01$ and $\tau=30.$}
\label{1Fig:11}\
\end{figure}

Figure \ref{1Fig:10} displays the effects of $\sigma$ (ion and electron temperature effect) in a collisionless (see upper two curves in Figure  \ref{1Fig:10})
as well as in a collisional plasma (see lower two curves in Figure  \ref{1Fig:10}). The considered plasma medium suggests the formation of smaller (in amplitude) and wider (in width) solitons in the presence of collision in comparison to those are formed in a collisionless plasma. It is also seen that for negligible temperature and collisional effects (i.e., $\sigma=0,\,\nu=0$), the solitary structure does not get damped hence the amplitude is maximum. On the other hand, in the presence of considerable amount of temperature and no collision, the structure gets slightly damped. For the reverse condition, i.e., with having collision in system but no temperature effect, the structure experiences damping comparatively on a large scale as that for the previous case. We noticed that in the presence of ion (electron) temperature effect and collisional effect (i.e., $\sigma\neq0,\,\nu\neq0$), the amplitude of the structure becomes minimum.

In Figure  \ref{1Fig:11}, we depict the DIASWs profile versus $\xi$ for different values of electron's superthermality  index $\kappa$ to trace the influence
of nonthermality on the dynamical properties of DIASWs in the considered plasma. We have observed another notable result from the numerical simulation, which shows that as the distribution of the plasma medium tends to go toward the thermal distribution from the superthermal one, the value of both amplitude and width of solitary waves increase.

\begin{figure}[!h]
\begin{center}
\includegraphics[width=6.7cm]{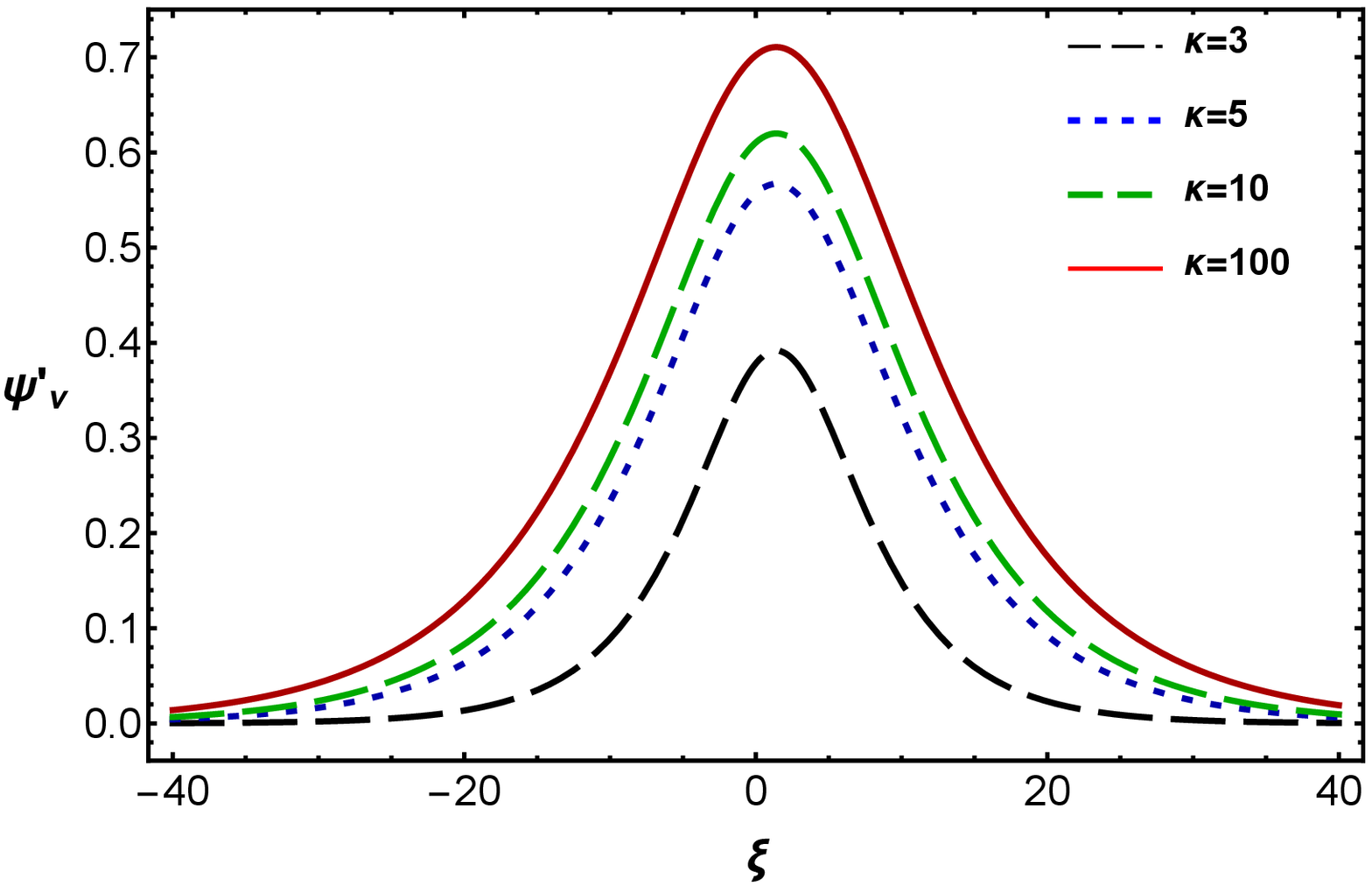}

\large{(a)}\vspace{0.6cm}

\includegraphics[width=6.7cm]{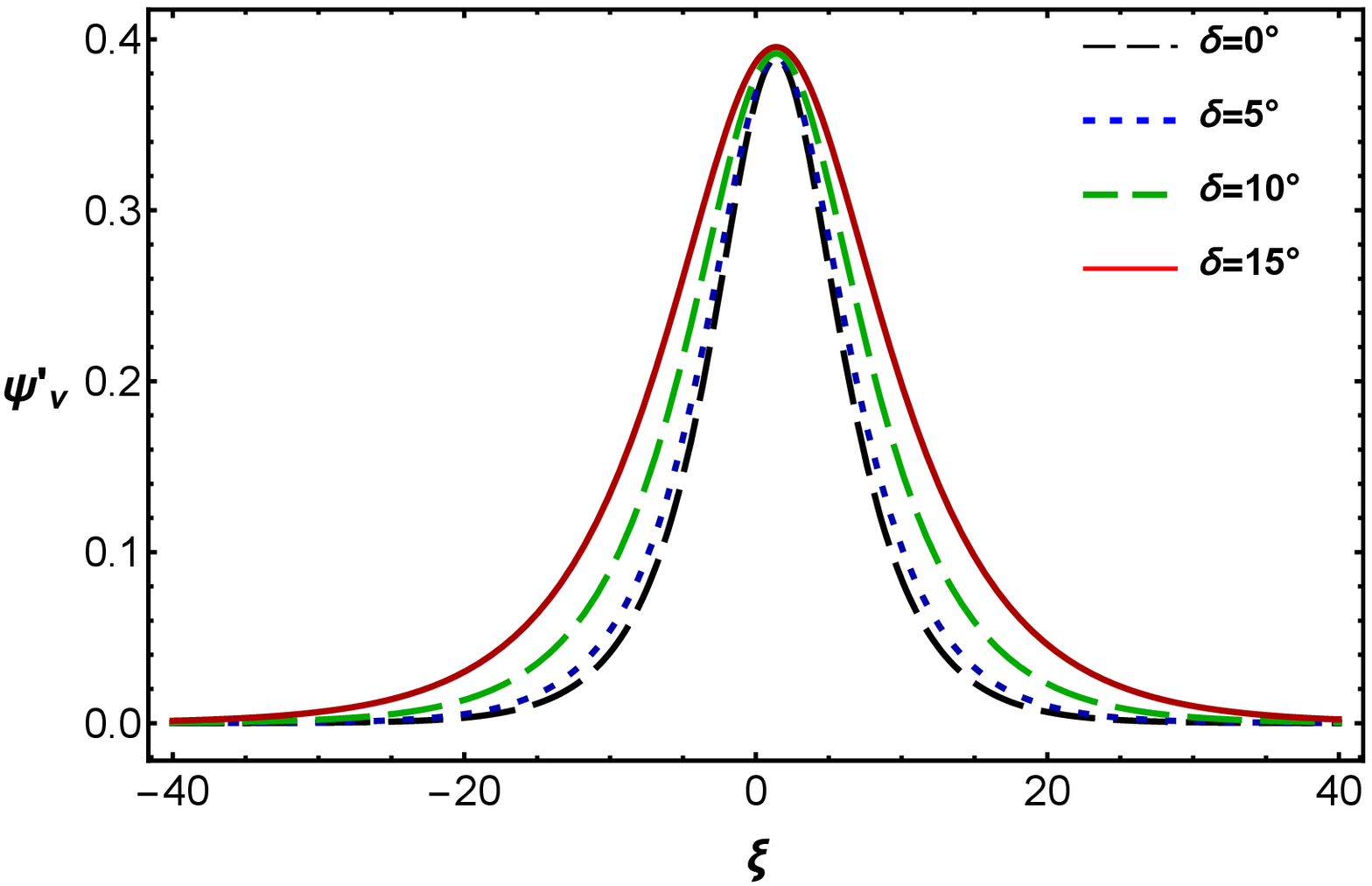}

\large{(b)}\vspace{0.6cm}

\includegraphics[width=6.7cm]{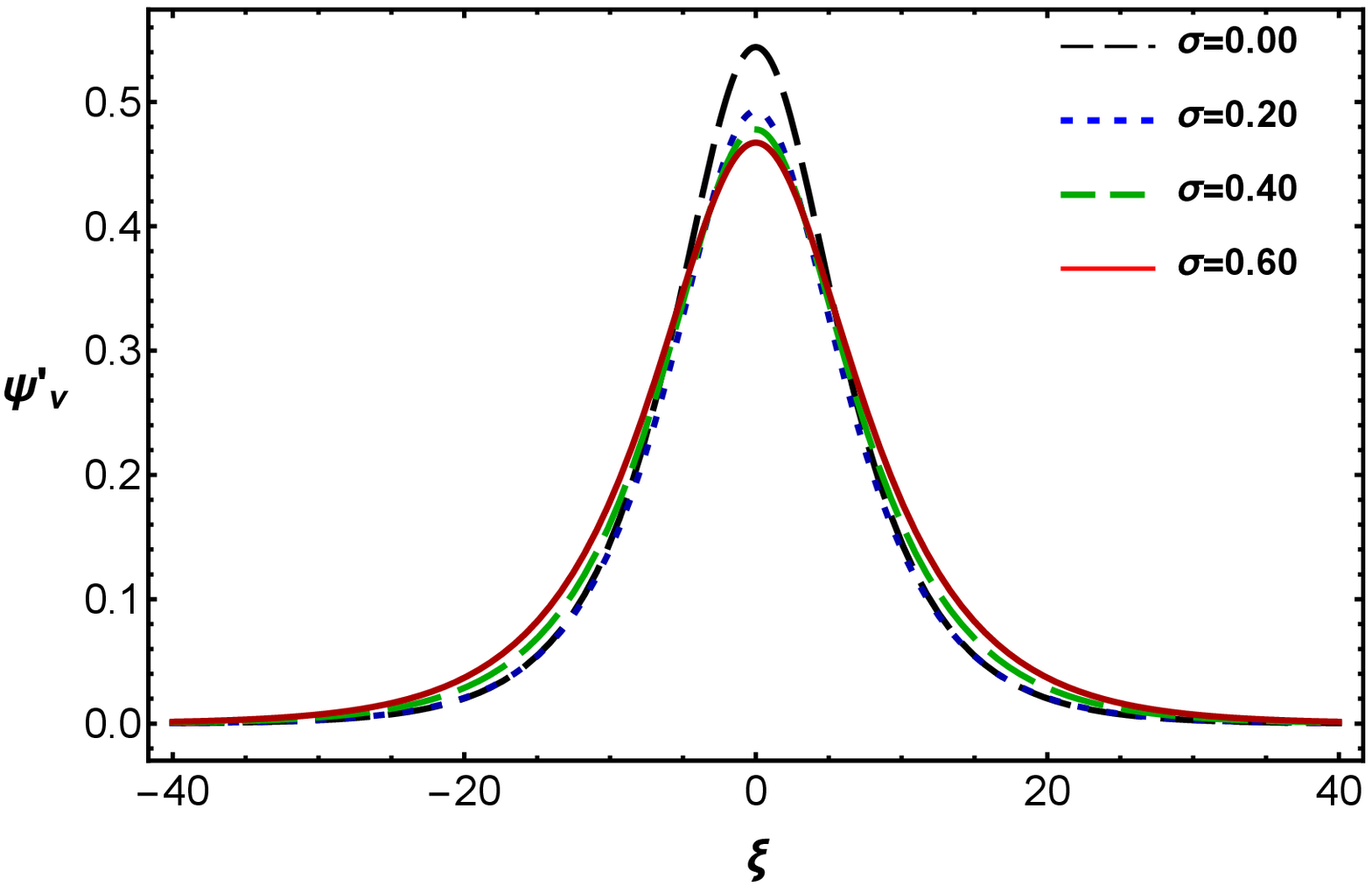}

\large{(c)}
\end{center}
\caption{Showing the variation of damped solitary potential [given in (\ref{1:eq24b})] (a) $\psi^\prime_{\nu}$ versus $\xi$ for different values of $\kappa$ with $\sigma=0.10$ and $\delta=10\si{\degree}$, (b) $\psi^\prime_{\nu}$ versus $\xi$ for different values $\delta$ with $\sigma=0.10$ and $\kappa=3$, and (c) $\psi^\prime_{\nu}$ versus $\xi$ for different values of $\sigma$ with $\delta=10\si{\degree}$ and $\kappa=3$, where other plasma compositional parameters are fixed at $\Omega_{ci}=0.2$, $\nu= 0.01$, $\tau=30$, $U=0.05$, and $\mu=\mu_c$.}
\label{1fig:12}\
\end{figure}

\textbf{The solitary excitation via the solitary wave solution of dmKdV equation for different plasma parameters are depicted in Figures \ref{1fig:12}.  It is predicted that the plasma nonthermality (via $\kappa$) has a significant role on the propagation nature of dmKdV solitary waves, and the amplitude  (width) of dmKdV solitons are seen to decrease (decrease) with the increase in plasma nonthermality, i.e., the taller and wider solitary waves are predicted to form in a Maxwellian plasma (higher $\kappa-$value) than those are formed in a nonthermal plasma (lower $\kappa-$value), as shown in Figure   \ref{1fig:12}a . It is eminent from Figure \ref{1fig:12}b that for the variation of obliqueness, even though the amplitude does not affect that much, the width gets influenced to a greater extent and gets increased with the increment of obliquity angle $\delta$. Finally, we examine the influence of ion temperature via $\sigma$ on the dmKdV solitary waves in Figure \ref{1fig:12}c, and we have investigated that the increase in the ion temperature leads to the formation of smaller and wider solitary waves in the considered plasma.}

\section{Conclusion}
\label{1sec:Discussion}
In this manuscript, we have  investigated the properties of DIASWs in a three component nonthermal (or superthermal) collisional magnetized dusty plasma medium (consisting of static dust particles, inertial ions, and inertialess electrons following nonthermal $\kappa-$distribution). The ion temperature and the ion-neutral collisional effects are taken into account. The reductive perturbation approach is adopted to derive the dKdV and dmKdV equation and the solitary wave  solutions of dKdV equation and dmKdV equations are obtained to model damped DIASWs and also to analyse their characteristics in such a plasma system. The influences of different  fundamental plasma parameters (e.g., obliquity angle, superthermality index, dust to ion ratio, magnetic field effect, etc.) in the presence of collision and ion temperature effects, are studied, and we have succeeded to extract some exciting results from our investigation which are briefly stated as follows:

\begin{enumerate}
  \item Ion temperature has a meaningful impact on the fundamental properties of DIASWs (e.g., stability, speed, amplitude, width, etc.) while advancing through a collisional magnetized dusty plasma.
  \item The Phase speed, $V$, is seen to be lesser while propagating obliquely (i.e., $0\si{\degree}<\delta<45\si{\degree}$) than while propagating along (i.e.,$\delta=0\si{\degree}$) the magnetic field. On the other hand, the DIAWs are predicted to propagate faster in Maxwellian plasmas than in superthermal (nonthermal) plasmas.
  \item The phase speed acquires a higher value in a dusty plasma with hot ion species than in a dusty plasma with cold ion species. It is also found that $V$ acquires higher value in a dusty plasma with more (less) population of dust (ions).
  \item The nonlinear coefficient $\alpha$ acquires both positive and negative value (depends on the concentration of dust and ion); thus, the propagation of ion-acoustic solitary structures of both positive and negative potential is possible in our considered plasma model. At a particular value of $\mu$, the nonlinearity coefficient vanishes (i.e., $\alpha=0$), which indicates that the amplitude of the solitary structure for this condition will be infinite.
  \item It is investigated that the ion temperature has a significant influence on the nonlinear term and is seen to increase with ion temperature, while the dispersion term does not suffer any mentionable impact caused by ion temperature.
  \item The dust ion-acoustic solitary waves suffer dissipation not only by the collision but also by ion temperature. However, the amount of dissipation caused by ion temperature drops down prominently in the presence of the ion-neutral collision.
  \item It is found that the magnetic field strength does not increase the amplitude of the solitary structures but contributes to the width to get broader.
  \item The ion temperature effect decreases the amplitude while increases the width of solitary waves, i.e., it plays the same role as the collision in this specific perspective, which is the same as we have found in some earlier works \cite{Mamun1997,Chatterjee2009,Roy2012}.
  \item If we consider both the ion temperature and the collision effect, the solitary wave width possesses a higher value in superthermal plasma comparatively than Maxwellian thermal plasma as expected. Nevertheless, surprisingly, the amplitude is also seen to be larger in superthermal plasma than in Maxwellian plasma, which is different from several existing published works \cite{El-Awady2010,Sing2013,Sultana2018} in several space and laboratory plasma studies.
  \textbf{\item At critical density $\mu_c$, the higher-order nonlinearity coefficient possesses positive values only. As a result, only compressive solitary waves propagate in our plasma medium.
  \item Both the amplitude and width of the dmKdV solitons suffer modifications with the change of different basic plasma parameters like superthermality index ($\kappa$), ion temperature effect ($\sigma$), and obliquity angle ($\delta$). With the increase of $\kappa$ and $\delta$ value of the system, both the amplitude and width of dmKdV solitons become bigger. However, the increase of ion temperature causes the dmKdV solitons to have smaller amplitude and more extended width.}

\end{enumerate}
Finally, we are expecting that this piece of investigative work will be handy to fathom more about the nonlinear structures and their basic properties that may originate and propagate in such a magnetized laboratory and space plasma medium where the electrons follow the superthermal distribution, the ions have a significant temperature which is not possible to disregard, and the collisional effect between ions and neutral particles plays a significant role.
\section*{DATA AVAILABILITY}
Research data are not shared.


\begin{thebibliography}{99}
\bibitem{Bliokh1985}P. V. Bliokh and V. V. Yarashenko, Sov. Astron. \textbf{29}, 330 (1985)
\bibitem{Shukla1992} P. K. Shukla and V. P. Silin, Phys. Scr. \textbf{45}, 508 (1992).
\bibitem{Rao1990} N. N. Rao, P. K. Shukla, and M. Y. Yu, Planet. Space Sci. \textbf{38}, 543 (1990).
\bibitem{Piel2002} A. Piel and A. Melzer, Plasma Phys. Controlled Fusion \textbf{44}, R1 (2002).
\bibitem{Temerin1982} M. Temerin, K. Cerny, W. Lotko, and F. S. Moze, Phys. Rev. Lett. \textbf{48}, 1175 (1982).
\bibitem{Ergun1998} R. E. Ergun, C. W. Carlson, J. P. McFadden et al., Geophys. Res. Lett. \textbf{25}, 2061 (1998).
\bibitem{Panwar2014} A. Panwar, C. M. Ryu, and A. S. Bains, Phys. Plasmas \textbf{21}, 122105 (2014).
\bibitem{Rehman2016} M. A. Rehman and M. K. Mishra, Phys. Plasmas \textbf{23}, 012302 (2016).
\bibitem{Chowdhury2017} N. A. Chowdhury, A. Mannan, M. M. Hasan, and  A. A. Mamun, Chaos \textbf{27}, 093105 (2017).
\bibitem{Chowdhury2018} N.A.Chowdhury, M.M.Hasan, A.Mannan, and A.A.Mamun, Vacuum \textbf{147}, 031 (2018).
\bibitem{Maksimovic1997} M. Maksimovic, V. Pierrard, and J.F. Lemaire, Astron. Astrophys. \textbf{324}, 725 (1997).
\bibitem{Nicholson1976} D. R. Nicholson and M. V. Goldman, Phys. Fluids. \textbf{19}, 1621 (1976).
\bibitem{Pereira1977} N. R. Pereira, Phys. Fluids. \textbf{20}, 1735 (1977).
\bibitem{Dutta2012} M. Dutta, S. Ghosh, and N. Chakrabarti, Phys. Rev. E \textbf{86} 066408 (2012).
\bibitem{Sultana2015} S. Sultana and I. Kourakis, Phys. Plasmas \textbf{22} (2015) 102302.
\bibitem{Mollenauer1980} L.F. Mollenauer, R.H. Stolen, and J.P. Gordon, Phys. Rev. Lett. \textbf{45} 1095 (1980).
\bibitem{Nozaki1983} K. Nozaki and N. Bekki, Phys. Rev. Lett. \textbf{51}, 2171 (1983).
\bibitem{Afanasjev1995} V. V. Afanasjev, Opt. Lett. \textbf{20}, 704 (1995).
\bibitem{Soto1997} J. M. Soto-Crespo et al., Phys. Rev. E \textbf{55}, 4783 (1997).
\bibitem{Chappell1980} C.R. Chappell and C. R. Baugher, Rev. Geophys. Space Phys. \textbf{18}, 853-861 (1980).
\bibitem{Vasyliunas1968} V. M. Vasyliunas and J. Geophys. Res. \textbf{73}, 2839 (1968).
\bibitem{Armstrong1983} T. P. Armstrong, M. T. Paonessa, E. V. Bell II, and S. M. Krimigis, J. Geophys. Res. \textbf{88}, 8893 (1983).
\bibitem{Fitzenreiter1998} F. J. Fitzenreiter and H. Abbasi, Geophys. Res. Lett. \textbf{25}, 249 (1998).
\bibitem{Vocks2008} C. Vocks, G. Mann, G. Rausche, Astron. Astrophys. \textbf{480}, 527 (2008).
\bibitem{Hasegawa1985} A. Hasegawa, K. Mima, and M. Duong-van, Phys. Rev. Lett. \textbf{54}, 2608 (1985).
\bibitem{Mace1995} R. L. Mace and M. A. Hellberg, Phys. Plasmas \textbf{2}, 2098 (1995).
\bibitem{Hellberg2009} M.A. Hellberg, R.L. Mace, T.K. Baluku, I. Kourakis, N.S. Saini, Phys. Plasmas \textbf{16}, 094701 (2009).
\bibitem{Abdelwahed2016} H. G. Abdelwahed, E. K. El-Shewy, M. A. Zahran, and S. A. Elwakil, Phys. Plasmas \textbf{23}, 022102 (2016).
\bibitem{Sing2019} K. Sing and N.S. Saini, Phys. Plasmas \textbf{26}, 113702 (2019).
\bibitem{Alinejad2019} H. Alinejad and M. Shahmansouri, IEEE Trans. Plasma Sci. \textbf{47}, 9 (2019).
\bibitem{Pakzad2011} H.R. Pakzad, Astrophys. Space Sci. \textbf{331}, 169 (2011).
\bibitem{Sultana2012} S. Sultana, G. Sarri, and I. Kourakis, Phys. Plasmas \textbf{19}, 012310 (2012).
\bibitem{Chahal2017} B.S. Chahal, Y. Ghai, and N.S. Saini, J. Theor. Appl. Phys. \textbf{11}, 181(2017).
\bibitem{Noman2019} A.A. Noman, N A Chowdhury, A. Mannan, and A. A. Mamun, Contrib. Plasma Phys. e201900023, (2019).
\bibitem{Sultana2011} S. Sultana, and I. Kourakis, Plasma Phys. Contr. Fusion. \textbf{53}, 045003 (2011).
\bibitem{Gill2010} T. S. Gill, C. Bedi, and A. S. Bains, Phys. Scr. \textbf{81}, 055503 (2010).
\bibitem{Saini2009} N. S. Saini, I. Kourakis, and M. A. Hellberg, Phys. Plasmas \textbf{16}, 062903 (2009).
\bibitem{Shahmansouri2013} M. Shahmansouri, Iranian J. Sci. Tech \textbf{37}, 285 (2013).
\bibitem{Farooq2017} M. Farooq, and M. Ahmad, Phys. Plasmas \textbf{24}, 123707 (2017).
\bibitem{Kunze1968} H.-J. Kunze, H.R. Griem, Phys. Rev. Lett. \textbf{21}  1048 (1968).
\bibitem{Ma1998} C. Ma and D. Summers, Geophys. Res. Lett. \textbf{25}, 4099 (1998).
\bibitem{Yoon2005} P. H. Yoon,T. Rhee, and C.-M. Ryu, Phys. Rev. Lett. \textbf{95}, 215003 (2005).
\bibitem{Naheer1978} E. Naheer, J. Hydraul. Res. 16, 235 (1978),
\bibitem{Vladimirov1999} S.V. Vladimirov, K.N. Ostrikov, and M.Y. Yu, Phys. Rev. E \textbf{60}, 3257 (1999).
\bibitem{Krapak2001} S.A. Krapak and G. Morfill, Phys. Plasmas \textbf{8}, 2629 (2001).
\bibitem{Mamun1997} A. A. Mamun, Phys. Rev. E \textbf{36}, 2 (1997).
\bibitem{Chatterjee2009} P. Chatterjee, K. Roy, S. V. Muniandy, S. L. Yap, and C. S. Wong, Phys. Plasmas \textbf{16}, 042311 (2009).
\bibitem{Roy2012} K. Roy, T. Saha, and P. Chatterjee, Phys. Plasmas \textbf{19}, 104502 (2012).
\bibitem{Mehdipoor2020} M. Mehdipoor, Wave Random Complex 1-25, (2020).
\bibitem{Wadati1975} M. Wadati, J. Phys. Soc. Jpn \textbf{38}, 673 (1975).
\bibitem{Kashkari2021} B. S. Kashkari and S. A. El-Tantawy, Eur. Phys. J. Plus \textbf{136}, 121 (2021).
\bibitem{Karpman1977} V. I. Karpman and E. M. Maslov, Sov. Phys. JETP \textbf{46}, 281 (1977).
\bibitem{Herman1990} R. L. Herman, J. Phys. A \textbf{23}, 2327 (1990).
\bibitem{Sahu2017} B. Sahu, A. Sinha and R. Roychoudhury, Phys. Plasmas \textbf{24}, 112111 (2017).
\bibitem{El-Awady2010} E.I. El-Awady, S.A. El-Tantawya, W.M. Moslem a, and P.K. Shukla, Phys. Lett. A \textbf{374}, 3216 (2010).
\bibitem{Sing2013} S. V. Singh, S. Devanandhan, G. S. Lakhina, and R. Bharuthram, Phys. Plasmas \textbf{20}, 012306 (2013).
\bibitem{Sultana2018} S.Sultana, Phys. Lett. A \textbf{382}, 1368 (2018).
\bibitem{Sultana2021} S Sultana, Chinese J. Phys. \textbf{69}, 206 (2021).
\bibitem{Mamun2021} A A Mamun and J Akter, J. Plasma Phys. 87, 905870109 (2021)
\bibitem{Thorne1991} R. M. Thorne and D. Summers, Phys. Fluids B \textbf{3}, 2117 (1991).
\bibitem{Singla2021} S. Singla and N.S.Saini, Results Phys. \textbf{22}, 103898 (2021).
\bibitem{Singh2006} D. K. Singh and H. K.Malik,  Phys. Plasmas \textbf{13}, 082104 (2006).
\end{thebibliography}
\end{document}